\UseRawInputEncoding
\documentclass[twocolumn,           
               showpacs,            
               nopreprintnumbers,     
               aps,                 
               prd,          	    
               letterpaper,             
              groupeaddress,      
               nofootinbib,         
               tightenlines,        
               floats,floatfix,      
               showkeys
               ]{revtex4-1}
               
\usepackage[toc,page]{appendix}
\usepackage{graphicx}
\usepackage{dcolumn}
\usepackage{bm}
\usepackage{amsmath}
\usepackage{amsfonts,amssymb}
\usepackage{soul}
\usepackage{color}
\definecolor{v}{rgb}{0.6, 0.2, 0.8} 
\definecolor{MAGA}{rgb}{0.1, 0.43, 0.75}
\definecolor{jm}{rgb}{0.13, 0.48, 0.64}

\usepackage{xcolor}
\usepackage{enumerate}
\usepackage{float}
\usepackage{subfigure}
\usepackage{multirow,tabularx, booktabs}
\usepackage{silence}
\usepackage{blindtext}
\begin{document}

\title{On a model of variable curvature that mimics the observed Universe acceleration}

\author{A. Esteban-Guti\'errez $^1$}
\email{ana.esteban@uv.cl}

\author{Miguel A. Garc\'ia-Aspeitia$^2$}
\email{angel.garcia@ibero.mx, corresponding author}

\author{A.  Hern\'andez-Almada$^3$}
\email{ahalmada@uaq.mx}

\author{Juan Maga\~na$^4$}
\email{juan.magana@ucentral.cl}

\author{V. Motta$^1$}
\email{veronica.motta@uv.cl}

\affiliation{$^1$Instituto de F\'isica y Astronom\'ia, Universidad de Valpara\'iso, Avda. Gran Breta\~na 1111, Valpara\'iso, Chile.}
\affiliation{$^2$ Depto. de F\'isica y Matem\'aticas, Universidad Iberoamericana Ciudad de M\'exico, Prolongaci\'on Paseo \\ de la Reforma 880, M\'exico D. F. 01219, M\'exico.}
\affiliation{$^3$ Facultad de Ingenier\'ia, Universidad Aut\'onoma de
Quer\'etaro, Centro Universitario Cerro de las Campanas, 76010, Santiago de 
Quer\'etaro, M\'exico}
\affiliation{$^4$Escuela de Ingenier\'ia, Universidad Central de Chile, Avenida Francisco de Aguirre 0405, 171-0164 La Serena, Coquimbo, Chile.}

\begin{abstract}
We present a new model based on General Relativity in where a subtle change of curvature at late times is able to produce the observed Universe acceleration and an oscillating behavior in the effective equation of state. This model aims to test the cosmological principle, by introducing a slight modification in the traditional FLRW metric, through a non-constant curvature parameter. This model is defined by a smooth step-like function with a slight transition between two curvature values, fulfilling the premise that the derivative of this curvature parameter is preserved as approximately zero, $\dot{\kappa}\approx0$. To test our model, we implemented a MCMC likelihood analysis using Cosmic Chronometers and Type Ia supernovae data in order to constrain the free parameters of the model and reconstruct $H(z)$, $q(z)$, $w_{eff}(z)$, also comparing the results with the $\Lambda$CDM model. The main result is that this model provides an alternative to the acceleration of the Universe without the need of a dark energy component. In particular, it gives an equivalent phase transition at $z \sim 0.5$, while obtaining the same fraction of matter density, similar to what is expected for the standard $\Lambda$CDM model. Remarkably, it also predicts a slight decelerated state at $z=0$ in agreement with diverse Dark Energy parameterizations. We conclude that the behavior of our proposed model points towards a new and intriguing way to investigate slight violations to the cosmological principle, in particular the case of inhomogenities during low phase transitions.
\end{abstract} 
\pacs{Dark energy, Variable curvature, Cosmology}
\maketitle

\section{Introduction}

Modern cosmology is based on General Relativity (GR), and from this the $\Lambda$-Cold Dark Matter ($\Lambda$CDM) model is constructed with a flat curvature. This model contains baryons, relativistic particles (radiation and neutrinos), dark matter (DM) and dark energy (DE) fluids, where, in particular, DE is considered as a cosmological constant. The success of the model is unprecedented and confirmed by several observations like the Cosmic Microwave Background Radiation (CMB) \cite{Planck:2018, Planck:2021}, Supernovae of the Ia Type (SNIa) \cite{Riess:1998,Perlmutter,Riess:2021jrx}, the Big Bang Nucleosynthesis (BBN) \cite{BBN-2017}, among others. However, the physical nature of the DM and the cosmological constant are still unknown, even though they comprise approximately the $\sim 95\%$ of our Universe. In particular, the cosmological constant is added \textit{ad-hoc} as a new fluid in the Friedmann equations with a constant equation of state (EoS) of $w=-1$ in order to have an accelerated Universe, and an energy density of $<10^{-10}$eV$^4$ to obtain a late acceleration ($z\sim 0.6$). However, if we accept the hypothesis of a cosmological constant that emerges from the quantum vacuum fluctuations, it is impossible to obtain the observed energy density, whose estimation differs by about $120$ orders of magnitude from the observed one \cite{Zeldovich,Weinberg}. This conundrum has generated several emergent DE models that, on one hand, aim to understand the Universe acceleration and, on the other hand, to extend the framework of GR in order to have new components that produce the observed de Sitter phase \cite{Motta:2021hvl,Amendola_2003,DeFelice:2010aj,Astorga-Moreno:2024okq}.
Despite the efforts in these two aspects to provide a more accurate explanation for the Universe acceleration, the cosmological constant is still the best candidate to understand this behavior. In summary, $\Lambda$CDM contains many afflictions, and as time passes, other issues are detected like the $H_0$ tension, CMB anisotropy anomalies, the age of the Universe, the universality of the cosmological principle, among others (see \cite{Perivolaropoulos:2021jda,Peebles:2024txt,LambdaCDMchallenges} for an excellent description of $\Lambda$CDM challenges).

On the other hand, several authors propose the exploration of curvature as a function dependent of the evolution of the Universe, i.e. $\Omega_\kappa(z)$. For example, in \cite{Liu:2020pfa}, the authors explore a $\Omega_\kappa(z)$ related to the Hubble parameter $H(z)$, constraining the curvature using quasar and SNIa data samples, obtaining $\Omega_\kappa=0.08\pm0.31$ and $\Omega_\kappa=-0.02\pm0.14$ respectively. Being both results also compatible with a flat Universe. In \cite{Liao:2017yeq}, authors propose a model-independent strategy by introducing a curvature parameter with $z$-dependence based on a linear Taylor expansion to test Friedmann-Lemaitre-Robertson-Walker (FLRW) line element. In particular, they test the hypothesis of homogeneity and isotropy (with a parameter characterizing the violation criterion) by taking the flat case, $\Omega_{\kappa}=0$, from the standard cosmological model as initial curvature. Their constrained results from simulations of time delays of gravitationally lensed quasars correspond to values of $\Omega_{\kappa}=0.057$ and $\Omega_{\kappa}=0.041$ for two different redshifts at $1\sigma$, which are consistent with the values found by \cite{Liu:2020pfa}. In particular, the authors also mention the necessity of a more robust theoretical study of $\kappa(z)$. 
Moreover, \cite{Rasanen:2014mca} also implement a curvature test, centering the study on geometric optics instead of focusing on the matter/energy contents as usual. Through their methods, the authors obtain the constraints $-1.22<\Omega_{0\kappa}<0.63$ and $-0.08<\Omega_{0\kappa}<0.97$ with a CMB prior and the local Hubble constant. In addition, in \cite{Heinesen-2020}, they proposed a modified FLRW equation characterized by a coupling between the curvature and structure formation, with the corresponding curvature function given by their Eq. (18). Meanwhile, other authors such as \cite{Godlowski-2004,Hashemi:2014lqa} provide a profound study of Stephani models \cite{Stephani:2003tm}, breaking the cosmological principle and proposing a specific form of the curvature function in order to resolve several cosmological problems. However, these kinds of models are also questioned by many authors regarding their observational validity \cite{Balcerzak-2015,Ong-2018}.

Thus, in this paper, we intend to explore a variable curvature (VC) model to see whether it is capable of accelerating the Universe within the GR framework, while avoiding the introduction of the cosmological constant in the equations. We stand our ideas in recent studies which resent doubts regarding the universal flatness assumption (e.g. \cite{Anselmi_2023}) or in where it is questioned if the Universe maintains the cosmological principle at all times. For instance, according to \cite{Aluri:2022hzs} the homogeneity and isotropy is not maintained when it is faced with recent observations like cosmic dipoles or galaxy spins, among others.
In our case of study, considering GR with a FLRW metric demands that the curvature parameter must satisfy $\dot{\kappa}=0$, where dot represents a time derivative. According to this restriction, we propose a function that slightly violates the cosmological principle only at late times in order to maintain the benefits of this symmetry. 
This function behaves as a step function moderated by a parameter $\gamma$, which plays the role of softening the transition between two curvatures. When the $\gamma$ parameter tends to infinity, we would retrieve a pure step function; however, when $\gamma$ is a small value, it takes on the role of the cosmological constant. Moreover, we add another parameter, $\alpha$, to obtain a variable curvature only within a certain redshift range, which we believe will happen in the local Universe without affecting the well-known primitive Universe, such as those during the primordial nucleosynthesis epoch. Additionally, we impose a change in curvature with a tendency to a hyperbolic Universe, while remaining close to flat curvatures, similar to the standard cosmological model, where currently: $\Omega_{0\kappa}^{\Lambda CDM}=0.001\pm0.002$ \cite{Planck:2018}, with $\Omega_{0\kappa}^{\Lambda CDM}=-\kappa^{\Lambda CDM}/H_0^2$. These requirements aim to satisfy the restriction $\dot{\kappa}\approx0$ and to be consistent within the FLRW framework, as we will see later in the discussion. 

Essentially, our proposed VC model is enclosed within a new framework that alters the Friedmann equation with the introduction of a curvature derivative, producing changes only within a specific redshift range during a curvature low-transition phase. Thus, the aim of this study is to show that, despite using GR, it is possible to modify the Friedmann equation and slightly break the cosmological principle in order to obtain a late acceleration without the need of a theory beyond GR (such as the \textit{f(R)-G} models; see e.g. other related works studying the role of the curvature in this context \cite{Granda-2020,Goswami-2023}). Is in this vein, we present the constraints of the VC model using Cosmic Chronometers, supernovae of Type Ia and a joint analysis of both as our data samples. Specifically, we will show that our results present interesting deviations, but are generally consistent with the standard cosmological paradigm. Remarkably, these differences could help us contrasting $\Lambda$CDM with alternative curvature variable models like the one presented in this paper, which could obtain a natural late-time acceleration.

The paper is outlined as follows: Section \ref{MF} presents and discusses the details of the mathematical formalism associated with a cosmology where the curvature is defined as a function of redshift, $\kappa(z)$, which varies between two close curvature values within a subtle phase transition. Sec. \ref{sec:constraints} is devoted to discuss the details of the observational samples used to constrain the free parameters of the model, where we mainly select local data samples. In Sec. \ref{sec:Res} we present our results and finally, in Sec. \ref{sec:DisandCon} we discuss the implications of the model with respect to the standard cosmology and other recent results on dark energy, concluding with our final remarks. We henceforth use units in which $c=\hbar=k_B=1$.

\section{Mathematical formalism} \label{MF}

We use a modified\footnote{This form has been thoroughly studied since first proposed by \cite{Krasinski:1981} as a modification of the RW metric. Since then, several studies have considered it as an effective cosmological model \cite{Paranjape:2008, Larena:2009}, discussed in the context of non-trivial spatial curvature for homogeneous and isotropic models in \cite{Rasanen:2008} and proposed as a possible GR generalization of a 
soluble non-relativistic cosmological model in \cite{Stichel:2018}.} FLRW-like line element with a time dependent curvature term $\kappa(t)$ as
\begin{eqnarray}
    ds^2=-dt^2+a(t)^2\left[\frac{dr^2}{1-\kappa(t)r^2}+r^2d\omega^2\right], \label{FLRWlinecurv}
\end{eqnarray}
where $d\omega^2 = d\theta^2+\sin^2\theta d\varphi^2$. Here, rather than modifying the time dependency with a deviation from a FLRW (such as the spherically symmetric inhomogeneous Stephani universes), we wanted to explore the idea of curvature remaining constant at most times, except for a small transition from one curvature to another at some redshift $z$. This transition must be smooth in order to only slightly violate the homogeneity presumption of the cosmological principle and continue using the benefits of the FLRW metric. Moreover, we also aimed to test whether that kind of transiting curvature behavior could mimic the effect of the cosmological constant in the acceleration of the Universe.

From the Einstein field equation $R_{\mu\nu}=8\pi G(T_{\mu\nu}-\frac{1}{2}g_{\mu\nu}T)$ and assuming a perfect fluid energy-momentum tensor $T_{\mu\nu}=pg_{\mu\nu}+(\rho+p)u_{\mu}u_{\nu}$, we solve the different terms of the equation, where $G$, $p$, $\rho$ and $u_{\mu}$ are the Newton gravitational constant, pressure, density and 4-velocity, respectively, in a comoving reference system. After some straightforward calculations in the Einstein tensor $tt$, $rr$ and $\theta\theta$ components (see Eqs. \eqref{uno}-\eqref{tres}), we arrive at the following modified Friedmann equation

\begin{eqnarray}
    H^2-\frac{H}{6}\frac{\dot{\kappa}}{\kappa}+\frac{\kappa}{a^2}=\frac{8\pi G}{3}\sum_i\rho_i, \label{FriedFinal}
\end{eqnarray}

\noindent
being $H\equiv\dot{a}/a$, where it is easy to see that it converges to the standard form when $\dot{\kappa}=0$. Notice that $\dot{\kappa}/\kappa$ does not diverge if $\kappa=0$, indeed we need to be careful and take this limit appropriately. For a flat curvature, we recover the standard Friedmann equation, assuming a constant curvature, $3H^2=8\pi G\sum_i\rho_i$. In this new Friedmann equation, we have an additional term related to a dynamical term associated with the curvature. However, the functional form of $\kappa$ will be restricted, as we will see later, in order to maintain the conditions of homogeneity and isotropy, at least partially.

Additionally, from the $tr$ component of Einstein equations, and according to a FLRW metric, we should need to fulfill that $r\dot{\kappa}(1-r^2\kappa)^{-1}=0$, which means returning to the case of constant curvature. However, we want to go beyond a constant curvature at all times, but approximately satisfying the cosmological principle. Thus, one way to perform such approximation to the previous equation (i.e. having $\dot{\kappa}\approx0$) is through the following form for curvature in terms of the redshift parameter
\begin{eqnarray}
 &&\kappa(z)=H_0^2\mathcal{H}(z),\label{kappa_function}
 \end{eqnarray}
 where 
\begin{eqnarray}
 &&\mathcal{H}(z)=\frac{\Omega^{1}_\kappa}{1+e^{-\gamma(z-\alpha)}}+\frac{\Omega^{2}_\kappa}{1+e^{\gamma(z-\alpha)}}=\nonumber\\&&\frac{1}{2}\Big\lbrace (\Omega_\kappa^1+\Omega_\kappa^2)+(\Delta \Omega_{\kappa})\tanh\left[\frac{\gamma(z-\alpha)}{2}\right]\Big\rbrace, \label{AnalyticCurv}
 \end{eqnarray}
where $\Delta \Omega_{\kappa} =\Omega_\kappa^1-\Omega_\kappa^2$. Additionally, $\dot{\kappa}$ takes the form
\begin{equation}
    \dot{\kappa}(z)=-(z+1)E(z)H_0^3\frac{d}{dz}\mathcal{H}(z),
\end{equation}
being $E(z)\equiv H(z)/H_0$. Notice that the $\dot{\kappa}$ function is constructed with units of $H_0^3$, thus, in terms of eV$^3$ it is expected to be small fulfilling the $rt$ equation. Also, both functions in Eq. (4) are equivalent and reproduce the same behavior, with the first one given in terms of the exponential function and the other in terms of the hyperbolic tangent function\footnote{Notice that the $\tanh$ form characterizes many emergent DE model such as \cite{PEDE:2019ApJ,Hernandez-Almada:2020uyr}.}. Note that $\mathcal{H}(z)$ simulates a step function (when $\gamma\to\infty$) where in this case, the parameter $\gamma$ is responsible for softening the transition, and we expect it to take the role of the parameter that triggers the acceleration of the Universe. The previous equation simulates the evolution of curvature, from $\Omega^{1}_\kappa$ in $z>\alpha$ to  
$\Omega^{2}_\kappa$ for $z\leqslant\alpha$, being $\Omega^{i}_\kappa$ our curvature density parameter defined as  $\Omega_\kappa^i\equiv \kappa^i/H_0^2$ \footnote{Here a difference from the standard notation arises, where a minus sign is normally introduced.} and is set as a free parameter, such as the $\alpha$ and the Hubble constant, $H_0$. In this context, the $\alpha$ parameter indicates the redshift at which the curvature transition takes place,  and it is expected to appear during the late times of the Universe's evolution. The previous Eq. \eqref{AnalyticCurv} satisfies $\dot{\kappa}\approx0$, when $\Omega_{\kappa}^i$ and $\gamma$ are small, provoking only a slight violation of the cosmological principle in the region of $z=\alpha$, allowing the initial prior on the cosmological principle hypothesis.

This can only be reached if we either demand small values for both curvature parameters and keep $\gamma$ small in order to have a smooth function between the two values of curvature, or if the transition occurs having a very large $\gamma$ in a very short time with our step function acting as a $\delta(z)$, regardless of the initial and final values of the curvature, thus providing an almost zero derivative in both cases. In fact, through a theoretical exploration of the parameters (see Appendix \ref{SubThe}), we found that among these two possible scenarios, the value of $\gamma$ must be small if we want to obtain a late acceleration as predicted by the current $\Lambda$CDM model, while larger values do not allow for the expected acceleration\footnote{Notice also, that despite the small number in eV of $\dot{\kappa}$, it is possible to obtain an accelerated expansion because the curvature transition is maintained during a long time in the last stages of the evolution of our Universe.}. 

On the other hand, the continuity equation for this model emerges when we demand $\nabla^{\mu}T_{\mu}^{\nu}=0$, thus we have

\begin{eqnarray}
\dot{\rho}+3H(\rho+p)=\frac{\dot{\kappa}}{2\kappa}(\rho+p). \label{Continuity}
\end{eqnarray}
Notice that a non-evolving constant curvature reproduces the traditional FLRW continuity equation. Assuming $p=w\rho$, where $w$ is the Equation of State (EoS), it is possible to integrate, having
\begin{equation}
    \rho=\rho_0\left(\frac{a_{0}}{a}\right)^{3(w+1)}\left(\frac{\kappa}{\kappa_0}\right)^{(w+1)/2}, \label{SolCE}
\end{equation}
where $a_{0}=1$, $\rho_0$ and $\kappa_0$ are appropriate integration constants,  and the function under the radical has an absolute value in order to obtain real values. In this case, it is important to adjust $\kappa_0$ in order to maintain a well-behaved density, which only has differences at $z=\alpha$. Therefore, we propose $\kappa_0=\xi H_0^2$, where $\xi$ is another tuned free parameter.

Combining Eqs. \eqref{FriedFinal} and \eqref{Continuity} and the corresponding derivatives, it is possible to obtain the acceleration equation as

\begin{eqnarray}
    &&\frac{\ddot{a}}{a}=\Big[2-\frac{\dot{\kappa}}{6H\kappa}\Big]^{-1}\Big\lbrace\frac{H}{6}\frac{\dot{\kappa}}{\kappa}+\frac{\ddot{\kappa}}{6\kappa}-\frac{\dot{\kappa}^2}{6\kappa^2}-\frac{\dot{\kappa}}{Ha^2}\nonumber\\&&+\frac{8\pi G}{3}\Big[\Big(\frac{\dot{\kappa}}{2\kappa H}-3\Big)(\rho+p)+2\rho\Big]\Big\rbrace.\label{accEq}
\end{eqnarray} 
Notice how Eq. \eqref{accEq} is now dependent on curvature derivatives, instead of the traditional acceleration equation. In this case, the corresponding acceleration of the Universe ($\ddot{a}>0$) is retrieved when the following differential equation is fulfilled

\begin{eqnarray}
    &&\ddot{\kappa}+\dot{\kappa}\Big[H-\frac{\dot{\kappa}}{\kappa}-\frac{6\kappa}{Ha^2}\Big]>-16\pi G\kappa\times\nonumber\\&&\Big[\Big(\frac{\dot{\kappa}}{2\kappa H}-3\Big)(\rho+p)+2\rho\Big].
\end{eqnarray}
Finally, after the theoretical tests described in Appendix \ref{SubThe}, we observe that the parameter $\xi$ from Eq. \eqref{SolCE} is related to $\Omega _{\kappa}^1$ in order to obtain the traditional behavior of perfect fluids, except when curvature is transitioning between $\Omega^{1}_\kappa$ and $\Omega^{2}_\kappa$, which occurs around $z\sim0$ (see Fig. \ref{fig:densityevol}). Thus, approximately, we have $\kappa_0\approx\Omega_{\kappa}^1H_0^2=\kappa$ at most times. Therefore, we can consider

\begin{equation}
    \rho_i\approx\rho_{i0}a^{-3(w_i+1)},\label{rho_approx}
\end{equation}
as a good approximation of a decoupled matter density field from the curvature term\footnote{As mentioned later in Section V, this choice is made to simplify the fluid treatment, so we can focus on the VC model as a candidate to avoid the DE inclusion causing the Universe acceleration. Nonetheless, this matter-curvature coupling during the transition needs to be further discussed to analyze its impact on local times.}. Thus, the dimensionless Friedmann equation is obtained by substituting Eq. \eqref{rho_approx} and the second definition of Eq. \eqref{AnalyticCurv} in Eq. \eqref{FriedFinal}. After that, calculating appropriately $\dot{\kappa}$ in terms of $z$, we finally have
\begin{eqnarray}
   &&E(z)^2\approx\left\lbrace1+\frac{\gamma(z+1)}{24\mathcal{H}(z)}(\Delta \Omega_{\kappa}) {\rm sech}^2\left[\frac{\gamma(z-\alpha)}{2}\right]\right\rbrace^{-1}\nonumber\\&&\times\left[\Omega_{m0}(z+1)^3 - \mathcal{H}(z)(z+1)^2\right], \label{eq:Final}
\end{eqnarray}
where $\Omega_{m0}$ is the matter density parameter at $z=0$ and we neglect the radiation component in this case, ($\Omega_{r0}\approx0$). Additionally, we have $a=(z+1)^{-1}$. The regions of the free parameters to explore are proposed under the premise of maintaining the Friedmann constraint $E(z=0)=1$. This form of $E(z)$ differs from the traditional one in the division given by the first term, which comes from the second term in Eq. \eqref{FriedFinal}. Thus, this term only comes into play during the subtle transition between curvatures, allowing us to recover the traditional dimensionless FLRW equation either before or after the transition.

On the other hand, the deceleration parameter can be constructed through the formula
\begin{eqnarray}
q(z)=\frac{(z+1)}{E(z)}\frac{d}{dz}E(z)-1, \label{DecPar}
\end{eqnarray}

where we have an accelerated Universe when $q<0$ and decelerated one when $q>0$. Moreover, the effective EoS is given by the equation

\begin{eqnarray}
w_{\mathrm{eff}}(z)=\frac{1}{3}[2q(z)-1],
\label{eq_weff}
\end{eqnarray}
where $q(z)$ is the deceleration parameter given by Eq. \eqref{DecPar}.

\section{Data and methodology} \label{sec:constraints}

The VC cosmology is confronted to Cosmic Chronometers (CC) and Type Ia supernovae (SNIa) datasets to constrain the phase-space with its free parameters given by ($h$, $\Omega_{m0}$, $\alpha$, $\gamma$, $\Omega^{1}_\kappa$, $\Omega^{2}_\kappa$) through a MCMC Likelihood analysis using \texttt{emcee} \cite{Foreman:2013} task in Python. We generate 4,000 chains with 250 steps each, after verifying their convergence using the auto-correlation function. 
The priors used are Gaussian distributions or $h = 0.7304 \pm 0.0104$ \cite{Riess:2019cxk} and $\Omega_{m0} = 0.3111\pm 0.0056$ \cite{Planck:2018}, and uniform distributions for $\Omega_\kappa^1\,\in[-0.1,0]$, $\Omega_\kappa^2\,\in[-0.1,0]$, $\alpha\,\in[-0.5,1]$ and $\gamma\,\in[1, 20]$. 
Stronger constraints are established by combining these two samples, which we will refer to as the joint analysis, with its $\chi^2$-function defined as
\begin{equation}\label{eq:chi2}
    \chi_{\rm Joint}^2 = \chi_{\rm CC}^2 + \chi_{\rm SNIa}^2 \,,
\end{equation}
where each term is the corresponding $\chi^2$ function per individual sample.

\subsection{Cosmic Chronometers}

Cosmic Chronometers is a dataset with 31 $H(z)$ measurements \cite{Moresco:2016mzx,Moresco:2015cya} in the redshift interval of $0.07<z<1.965$, based on the differential age (DA) strategy, which makes it an independent cosmological probe. Since these observations are considered uncorrelated, the  $\chi^2$ function can be expressed as
\begin{equation} \label{eq:chiOHD}
    \chi^2_{{\rm CC}}=\sum_{i=1}^{31}\left(\frac{H_{th}(z_i)-H_{obs}(z_i)}{\sigma^i_{obs}}\right)^2,
\end{equation}
where the $H_{obs}(z_i) \pm \sigma_i$ term represents the measurements of the Hubble parameter at the redshift $z_i$ and its $68\%$ confidence level uncertainty. The theoretical counterpart is represented as $H_{th}(z_i)$, estimated using our Eq. \eqref{eq:Final}.

\subsection{Type Ia supernovae (Pantheon+)}

Pantheon+ \cite{Scolnic2018-qf, Brout_2022} is the largest sample of SNIa observations,  corresponding to 1,701 distance modulus measurements spanning throughout the redshift range $0.001<z<2.26$. Since these measurements come from 1,550 different SNIa, the $\chi^2$ function is built \cite{Conley2010} as 
\begin{equation}\label{eq:chi2SnIa}
    \chi_{\rm SNIa}^{2}=a +\log \left( \frac{e}{2\pi} \right)-\frac{b^{2}}{e},
\end{equation}
where
\begin{eqnarray}
    a &=& \Delta\boldsymbol{\tilde{\mu}}^{T}\cdot\mathbf{Cov_{P}^{-1}}\cdot\Delta\boldsymbol{\tilde{\mu}}, \nonumber\\
    b &=& \Delta\boldsymbol{\tilde{\mu}}^{T}\cdot\mathbf{Cov_{P}^{-1}}\cdot\Delta\mathbf{1}, \\
    e &=& \Delta\mathbf{1}^{T}\cdot\mathbf{Cov_{P}^{-1}}\cdot\Delta\mathbf{1}\, \nonumber
\end{eqnarray}
and $\Delta\boldsymbol{\tilde{\mu}}$ defined as the difference vector between the observed distance modulus and the theoretical estimates given by 
\begin{equation}
    m_{th}=\mathcal{M}+5\log_{10}\left[\frac{d_L(z)}{10\, {\rm pc}}\right],
\end{equation}
where $\mathcal{M}$ is a nuisance parameter which is marginalized in the Eq. \eqref{eq:chi2SnIa} and the luminosity distance ($d_{L}$) is chosen for a hyperbolic Universe (covering also a flat case as we can see in Eq. \eqref{limithypflat}):

\begin{equation}\label{eq:dL}
   d_L(z)=(1+z)\frac{c}{H_0\sqrt{\Omega_{\kappa}}} \sinh\left[\sqrt{\Omega_{\kappa}}\int_0^z \frac{dz^{\prime}}{E(z^{\prime})}\right] \,,
\end{equation}
being $c$ the speed of light and $H_0$ the Hubble constant. Additionally $\Delta\mathbf{1}=(1,1,\dots,1)^T$ is
the transpose of the vectors
and $\mathbf{Cov_{P}}$ is the covariance matrix, which includes both systematic and statistic uncertainties.

 \section{Results} \label{sec:Res}

We present the results of our statistical likelihood analysis, applying the form of $E(z)$ described in Eq. \eqref{eq:Final}, and proposing a particular behavior focused on a small curvature transition given by Eq. \eqref{AnalyticCurv} and satisfying the curvature derivative is approximately zero at all times. As already commented in Section II, this is plausible only because we assume the values of $\Omega_{\kappa}^i$ and $\gamma$ to be small, according to the priors discussed in Section III.

\begin{figure*}
    \centering
    \includegraphics[width=0.65
\textwidth]{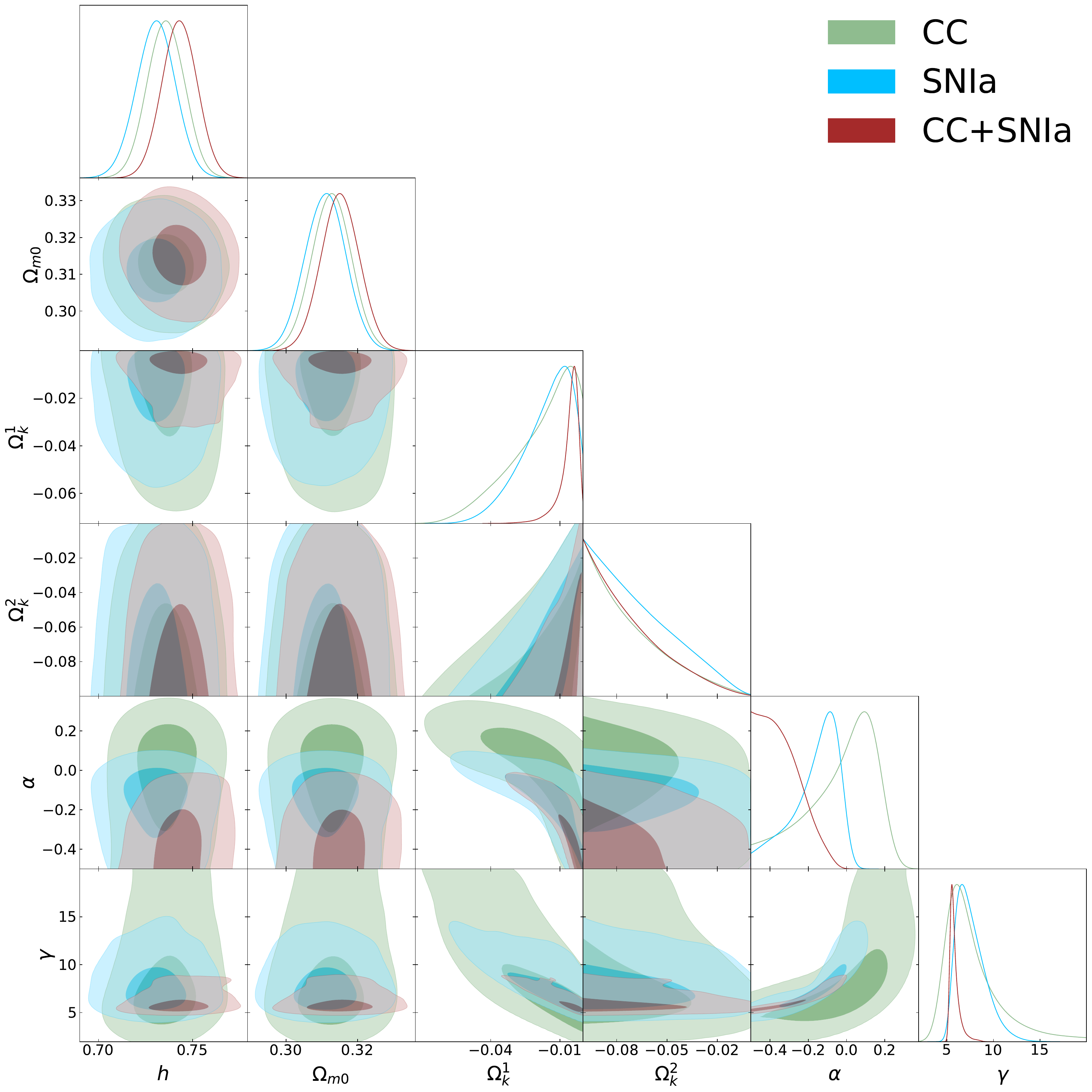}
    \includegraphics[width=0.65\textwidth]{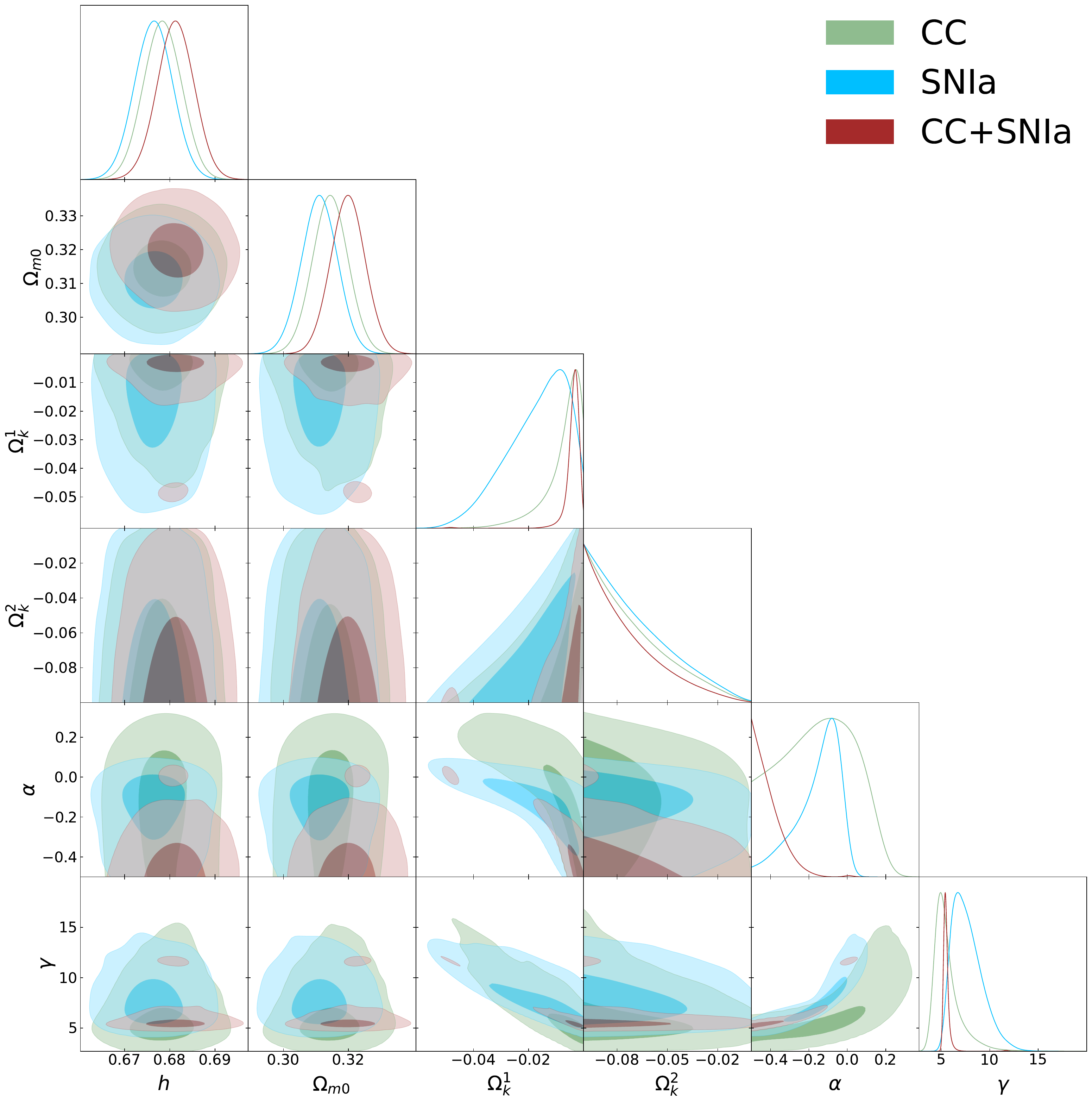}
    \caption{1D posterior distributions and 2D contours at $1\sigma$ (inner region) and $2\sigma$ (outermost region) CL for the VC model are shown.  Top (bottom) panel is displayed contours with Gaussian prior on SH0ES (Planck) $H_0$ value. The green, blue, and red PDFs correspond to constraints from CC, SNIa and CC+SNIa, respectively.}
    \label{fig:contoursSH0ES}
\end{figure*}

\begin{figure*}
    \centering
    \includegraphics[width=0.45\textwidth]{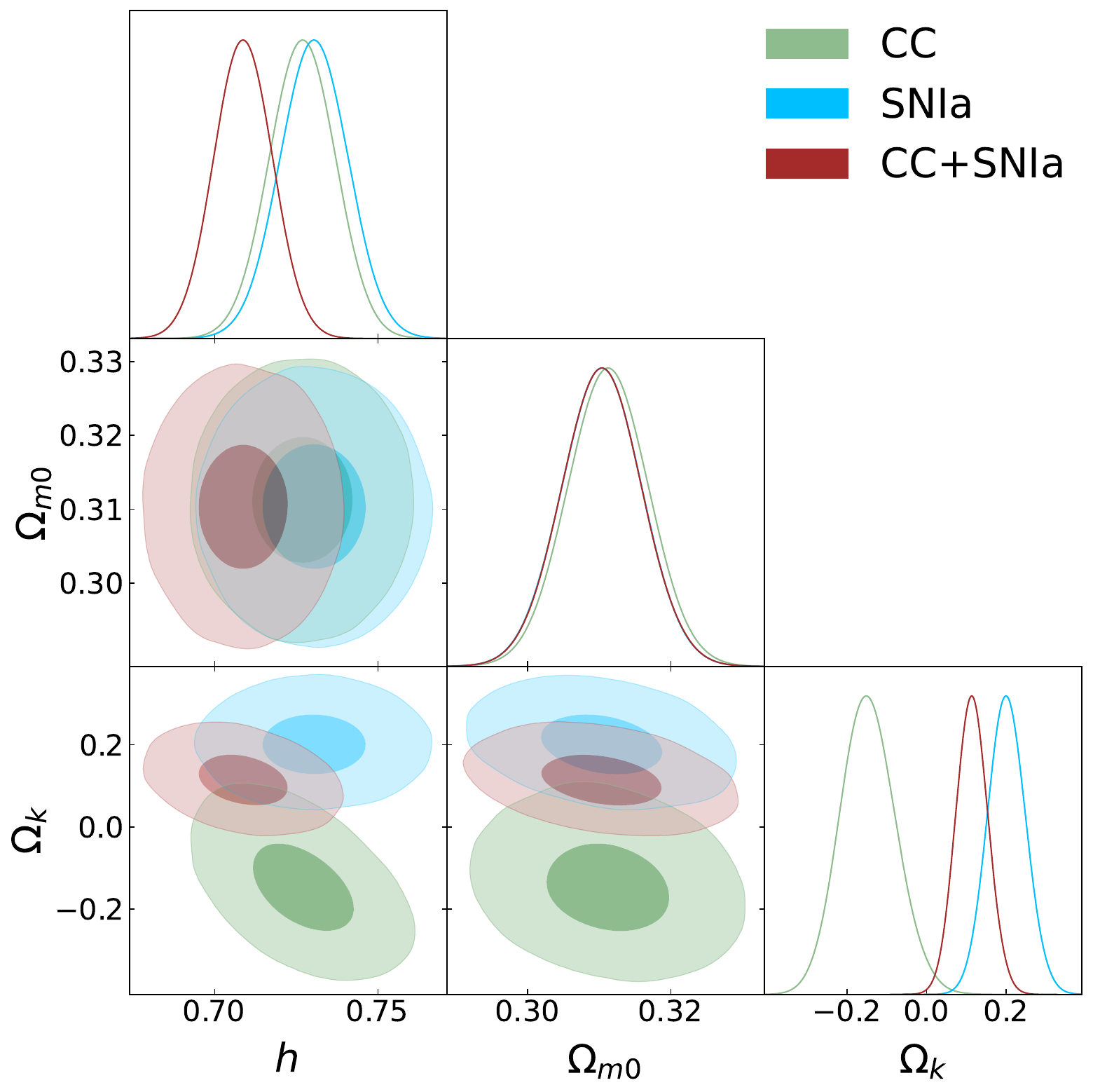} 
    \includegraphics[width=0.45\textwidth]{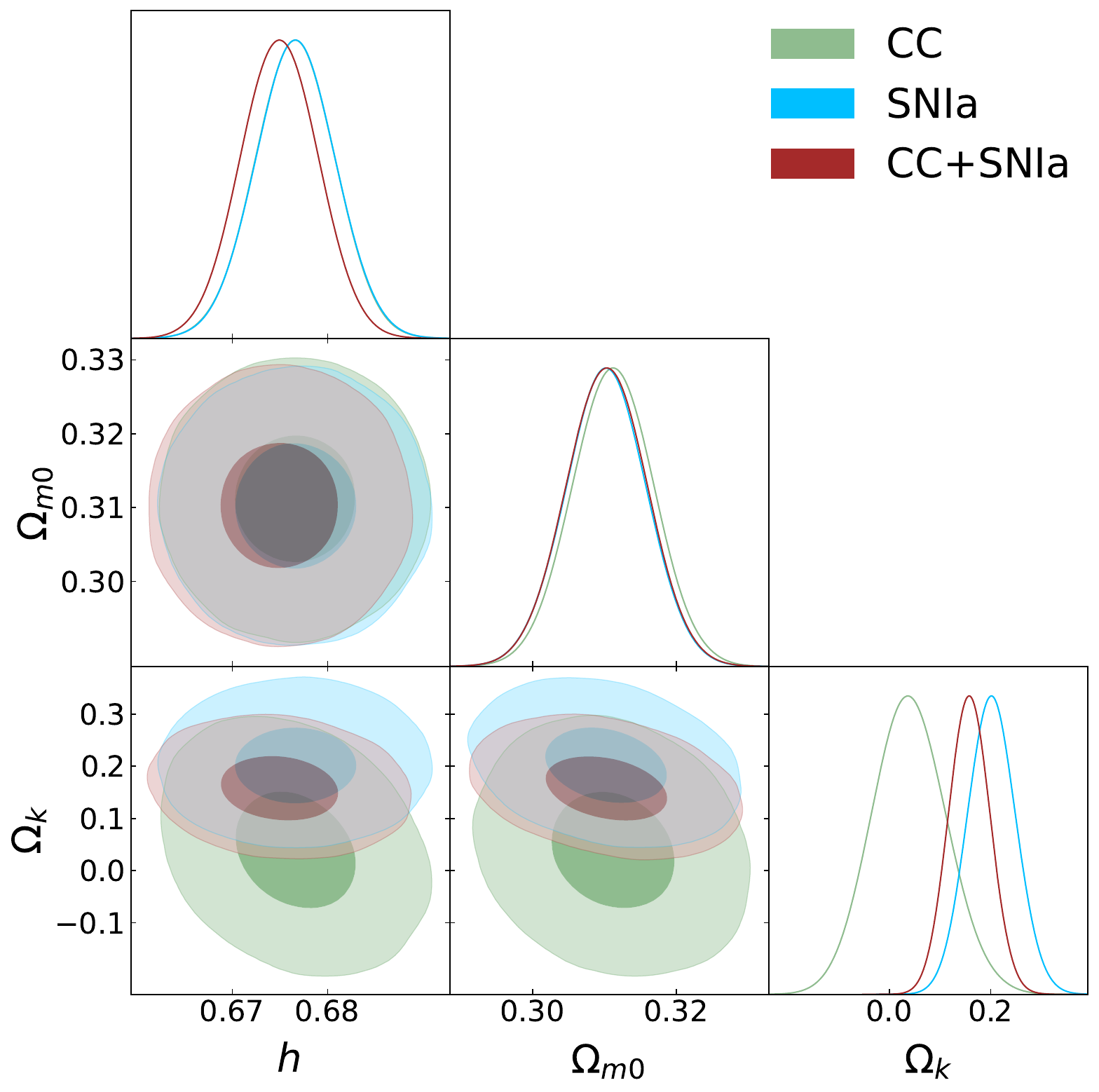} 
    \caption{1D posterior distributions and 2D contours at $1\sigma$ (inner region) and $2\sigma$ (outermost region) CL for the $\Omega_{\kappa}$-$\Lambda$CDM model are shown. Left (right) panel is displayed contours with Gaussian prior on SH0ES (Planck) $H_0$ value. The green, blue, and red PDFs correspond to constraints from CC, SNIa and CC+SNIa, respectively.}
    \label{fig:contoursPlanck}
\end{figure*}

Our marginalized values, along with the corresponding 2D contours at the 68\% ($1\sigma$) and 95\% ($2\sigma$) confidence level and the 1D posterior distributions, are presented in Fig. \ref{fig:contoursSH0ES} for the parameters of the VC model using CC, SNIa, and the joint analysis CC+SNIa with SH0ES (top panel)  and Planck priors (bottom panel). For comparison purposes, we also tested the data with the same initial priors for the standard $\Lambda$CDM model with free curvature $\Omega_k$, with the corresponding results shown in Fig. \ref{fig:contoursPlanck}. Table \ref{tab:bf_model} presents the intervals of the parameters at $1\sigma$ confidence level for both the VC model and $\Omega_k$-$\Lambda$CDM with some upper and lower limits estimated at a 90\% confidence level. The quality of the fit is estimated by computing the reduced-$\chi^2$ as $\chi^2_{red}=\chi^2/{\rm ndf}$ for CC, SNIa, and CC+SNIa,  where ndf is the difference between the size of the sample and the number of free model parameters.
The constraints preserve the priors established over the dimensionless Hubble parameter ($h$) on the SH0ES and Planck values \citep{Riess:2021jrx, Planck:2018} and for the matter density component ($\Omega_{m0}$) corresponding to the latest Planck result \cite{Planck:2018}. As previously commented, the curvature phase transition is restricted to vary in the negative region ($-0.1<\Omega_\kappa^i<0$) to remain close to the equality $E(0)=1$ and produce an accelerated stage. However, depending on the combination of the model's free parameters, this value may shift by a small amount. Finally, according to our constraints, using the parameter distribution of the marginalized values for the different samples with respect to $E(0)=1$, and taking into account the uncertainty propagation with respect to the central value of $E(0)$, we obtain deviations in $E(0)$ around $17\%-22\%$. In this context, the value of $H_0$ must be recalculated as $H(0)= 78\%-82\% H_0$  and $H(0)= 77\%-85\% H_0$ by imposing the SH0ES and Planck priors, respectively. From here we first see that our results show an overall reduction between 23$\%$ and 15$\%$ for both priors. On the one side, if we take the SH0ES value of $H_0$, our VC model may help to alleviate the Hubble tension paradigm by reducing its value to a fewer percentage. On the other side, since this change also appears for the Planck prior, we cannot state those values as conclusive, yet a more detailed characterization of the initial priors within the VC scenario plus more observational constraints for the late Universe are needed to either confirm or reject this result. 

\begin{figure*}
    \centering
    \includegraphics[width=0.32\textwidth]{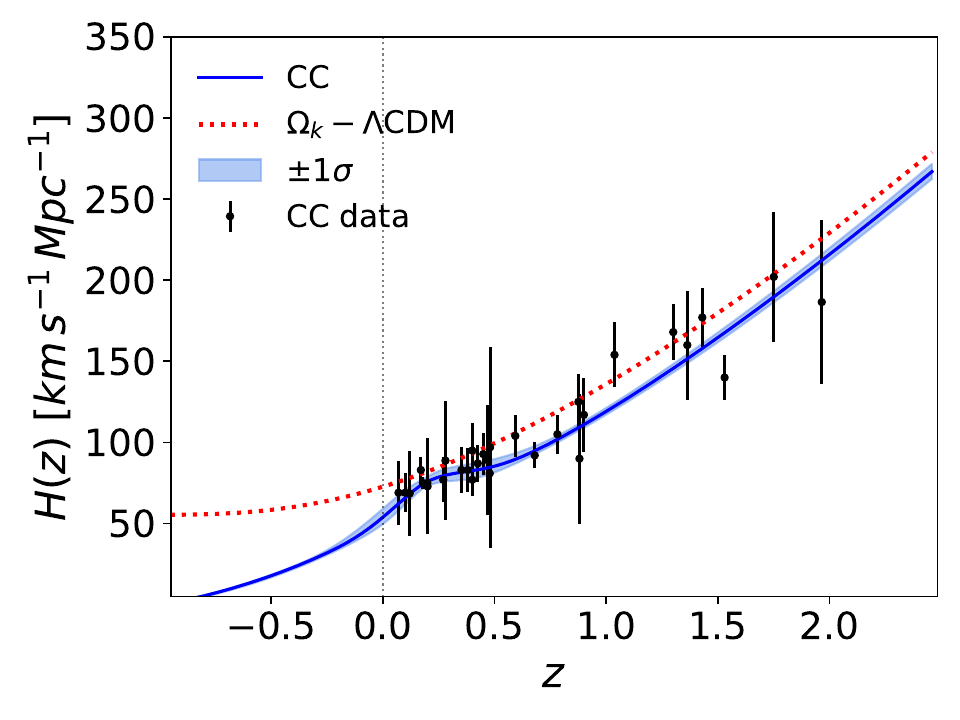}
    \includegraphics[width=0.32\textwidth]{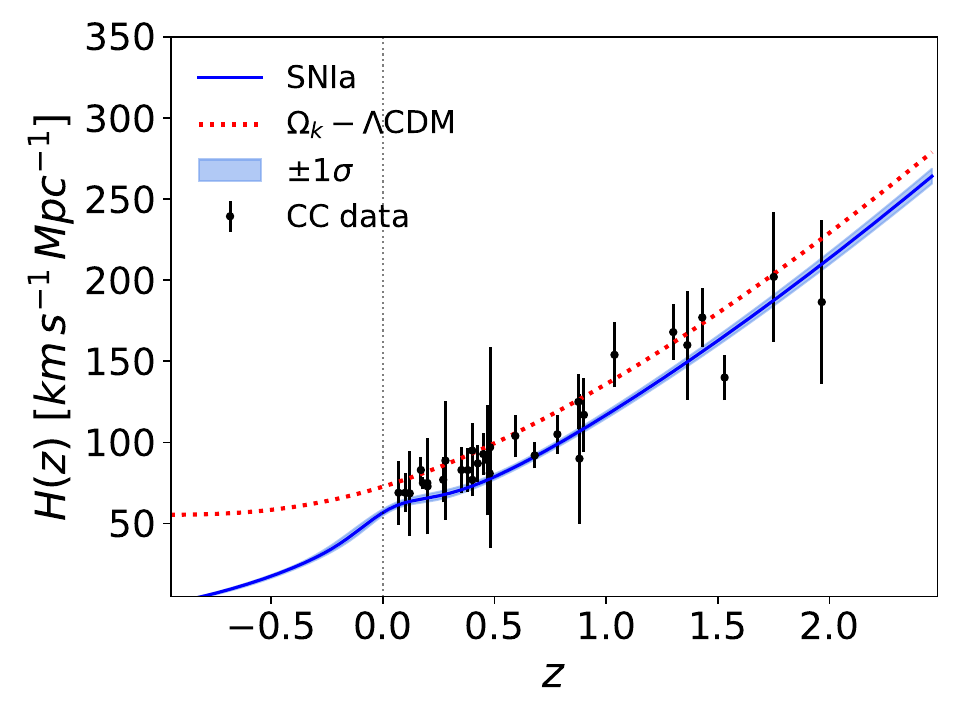}
    \includegraphics[width=0.32\textwidth]{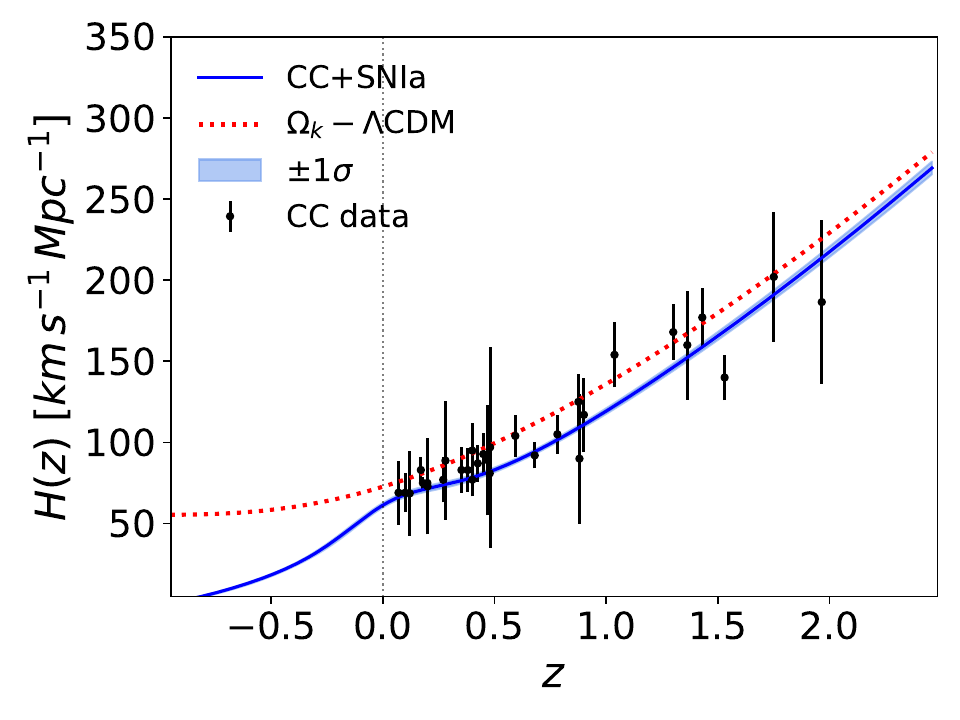}\\
    \includegraphics[width=0.32\textwidth]{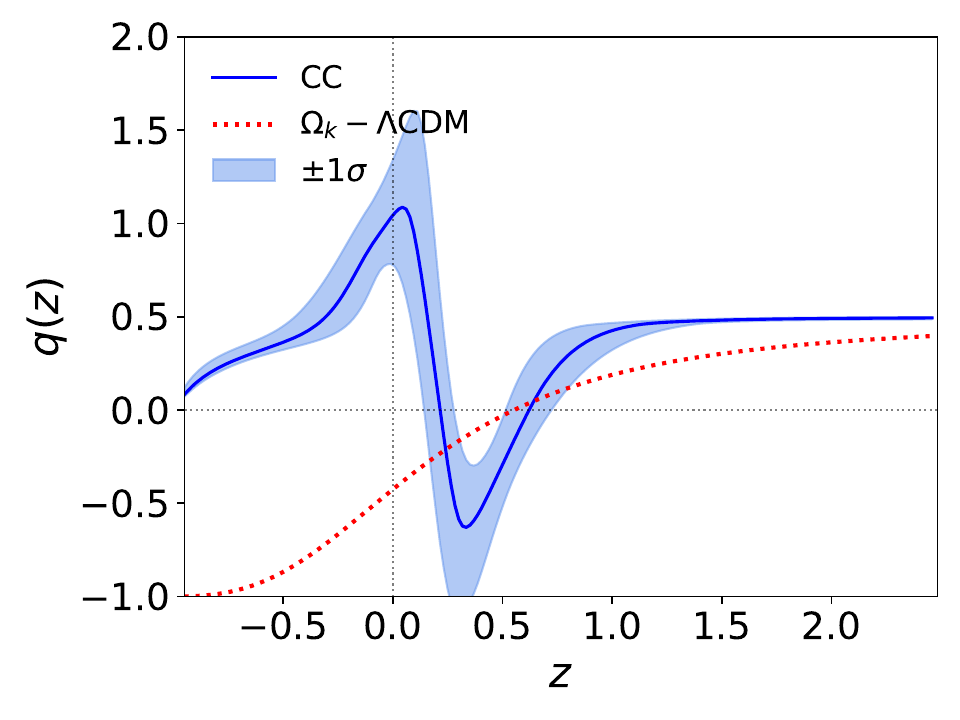}
    \includegraphics[width=0.32\textwidth]{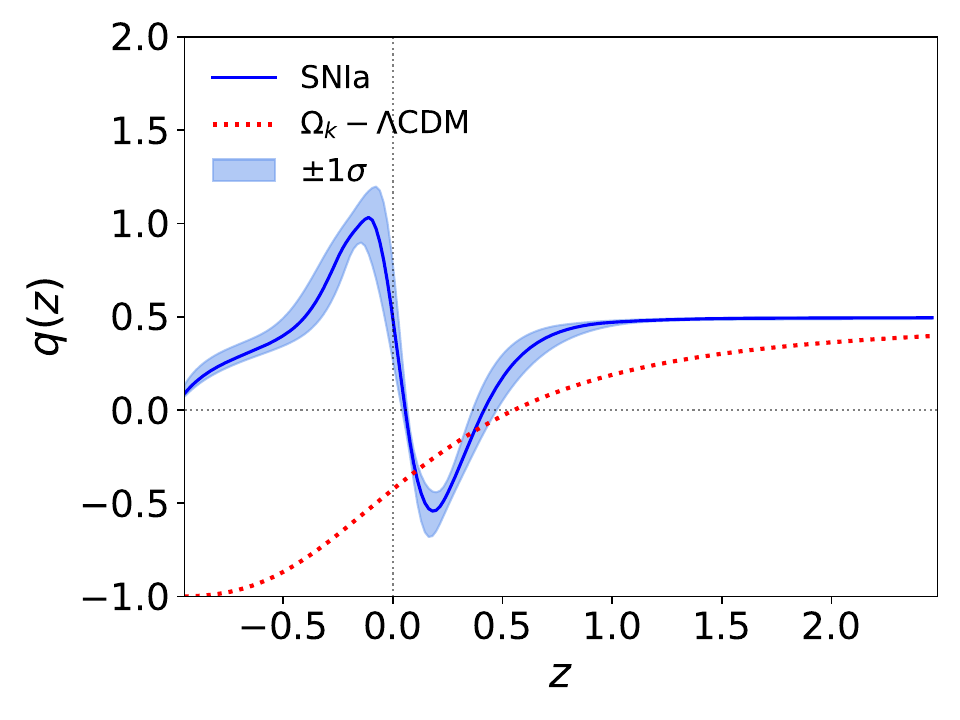}
    \includegraphics[width=0.32\textwidth]{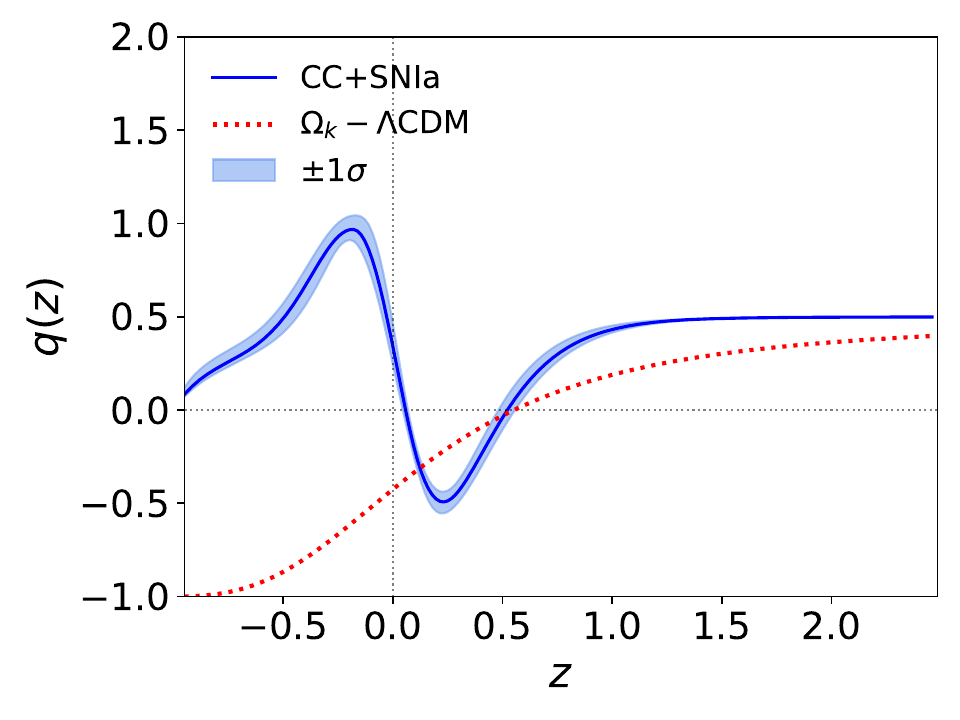}\\  
    \includegraphics[width=0.32\textwidth]{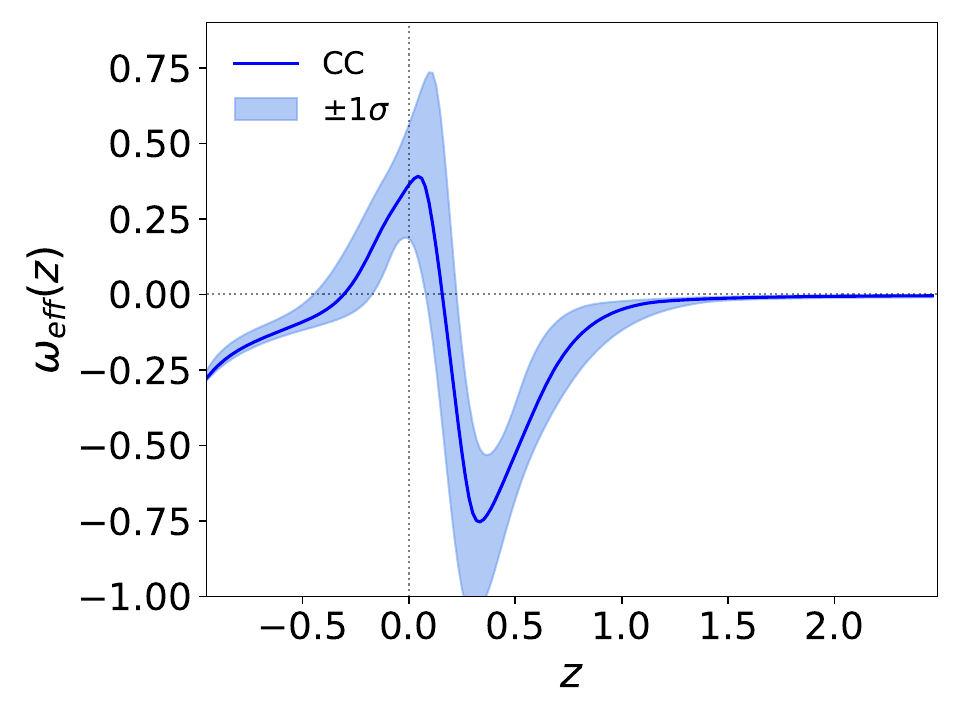}
    \includegraphics[width=0.32\textwidth]{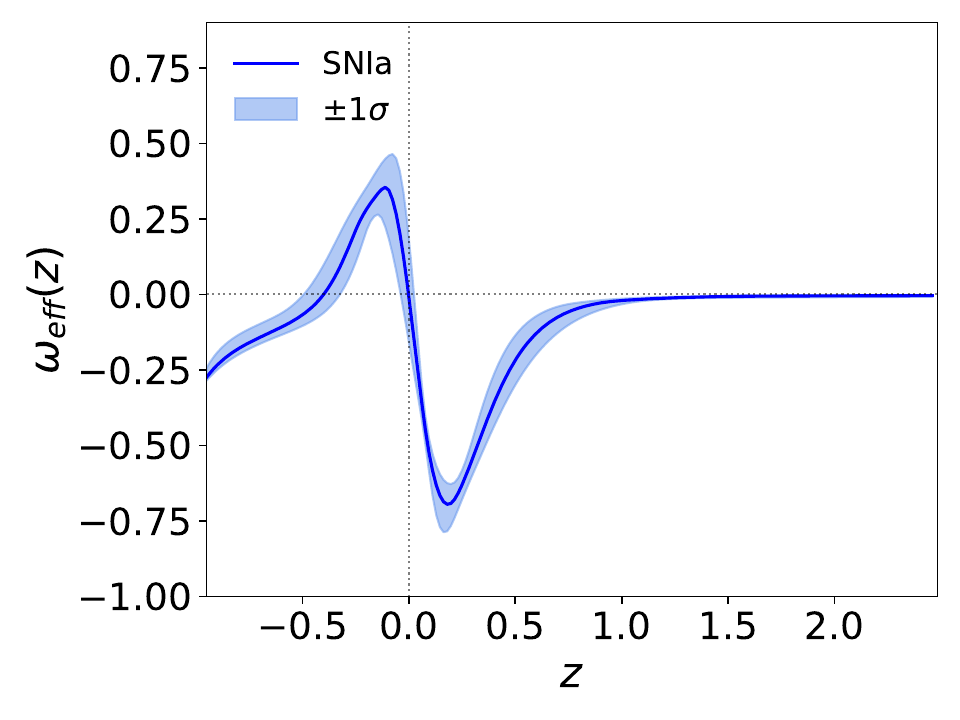}
    \includegraphics[width=0.32\textwidth]{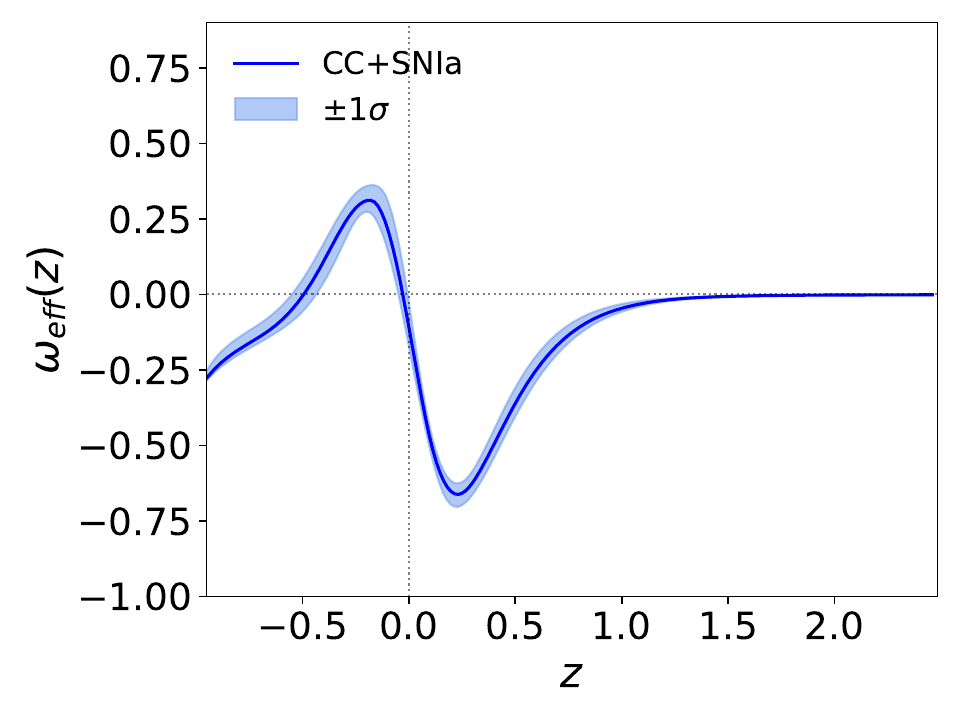}\\ 
    \includegraphics[width=0.32\textwidth]{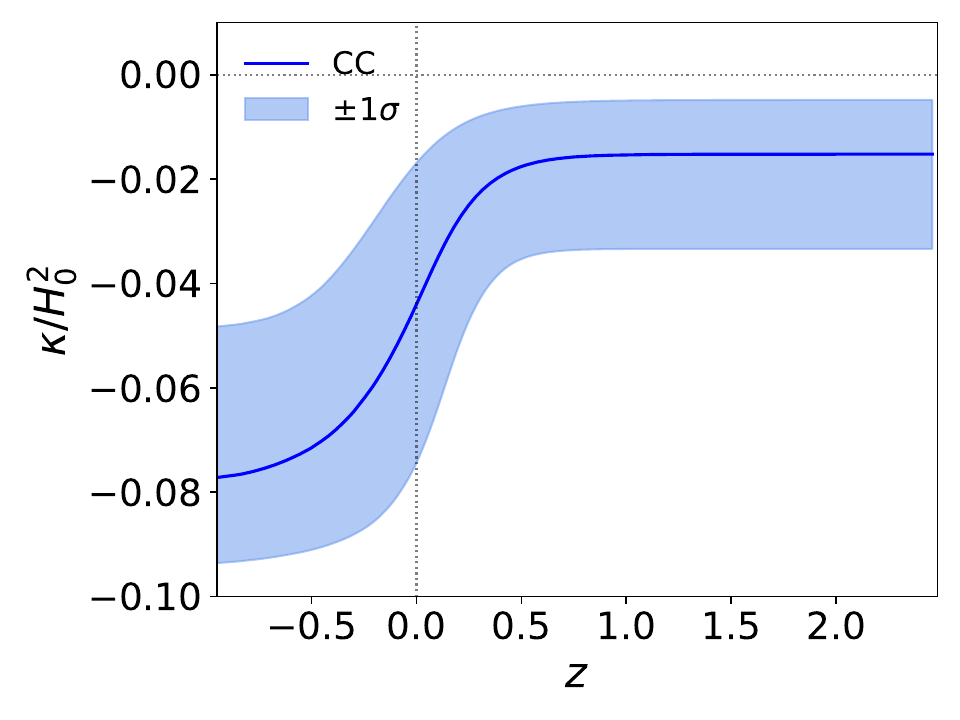}
    \includegraphics[width=0.32\textwidth]{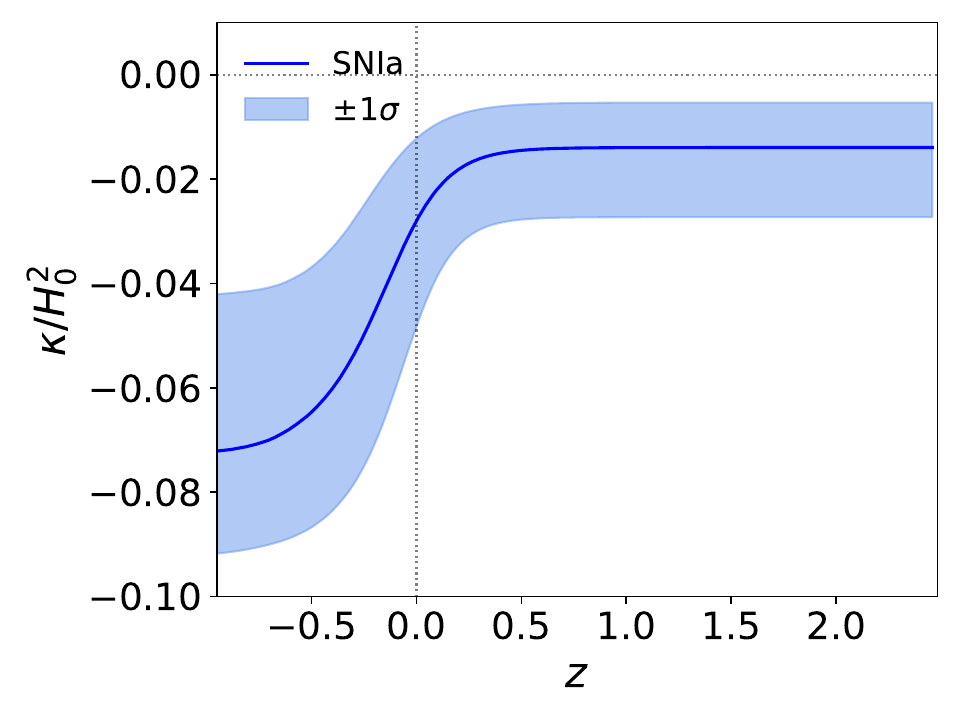}
    \includegraphics[width=0.32\textwidth]{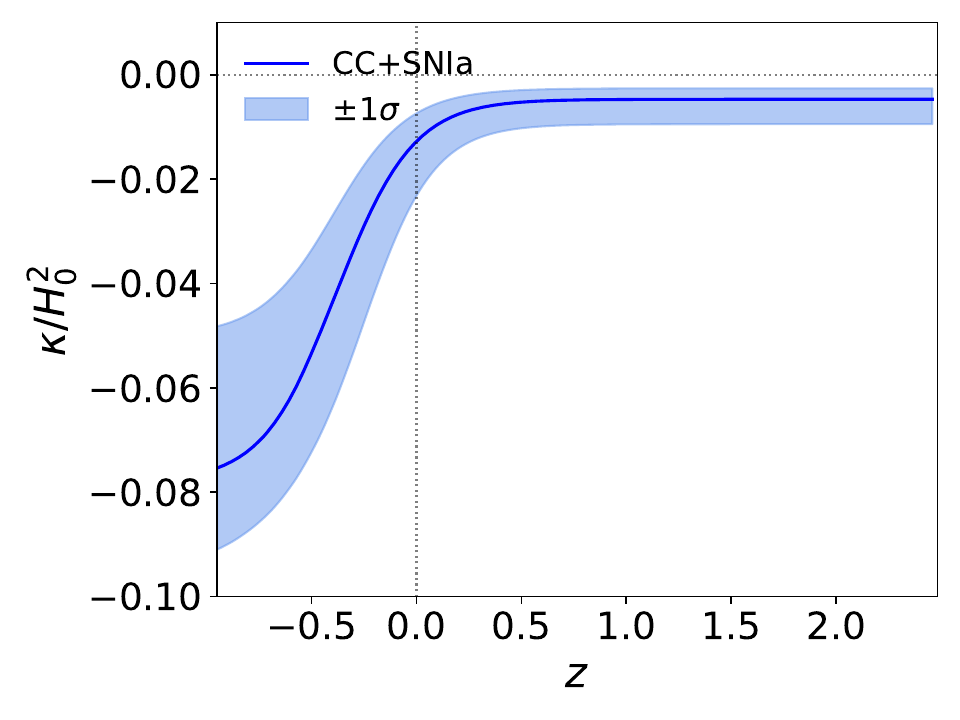}\\ 
    \caption{Reconstruction of the Hubble parameter (first panel), the deceleration parameter (second panel), the effective EoS (third panel) and curvature (fourth panel) for the VC model in the redshift range $-1<z<2.2$ using CC (left), SNIa (middle) and CC+SNIa (right) datasets taking into account Gaussian prior on SH0ES $H_0$ value. The bands represent $1\sigma$ CL uncertainties around the best-fit values (solid blue line). The standard $\Lambda$CDM model is included in the two first panels as red dashed lines.}
    \label{fig:cosmographySH0ES}
\end{figure*}

\begin{figure*}
    \centering
    \includegraphics[width=0.32\textwidth]{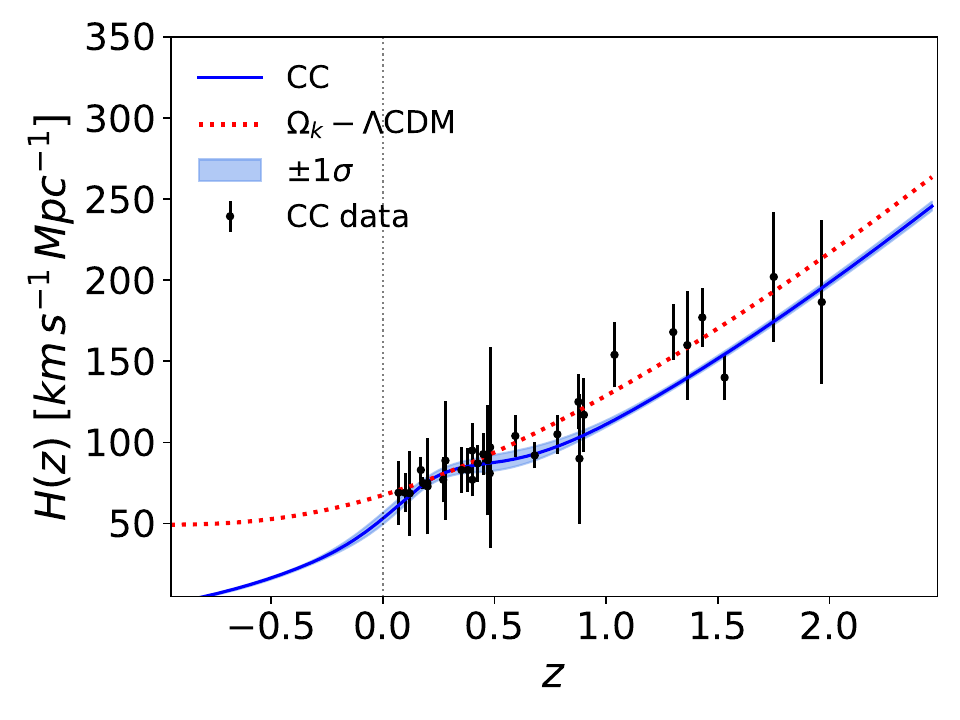}
    \includegraphics[width=0.32\textwidth]{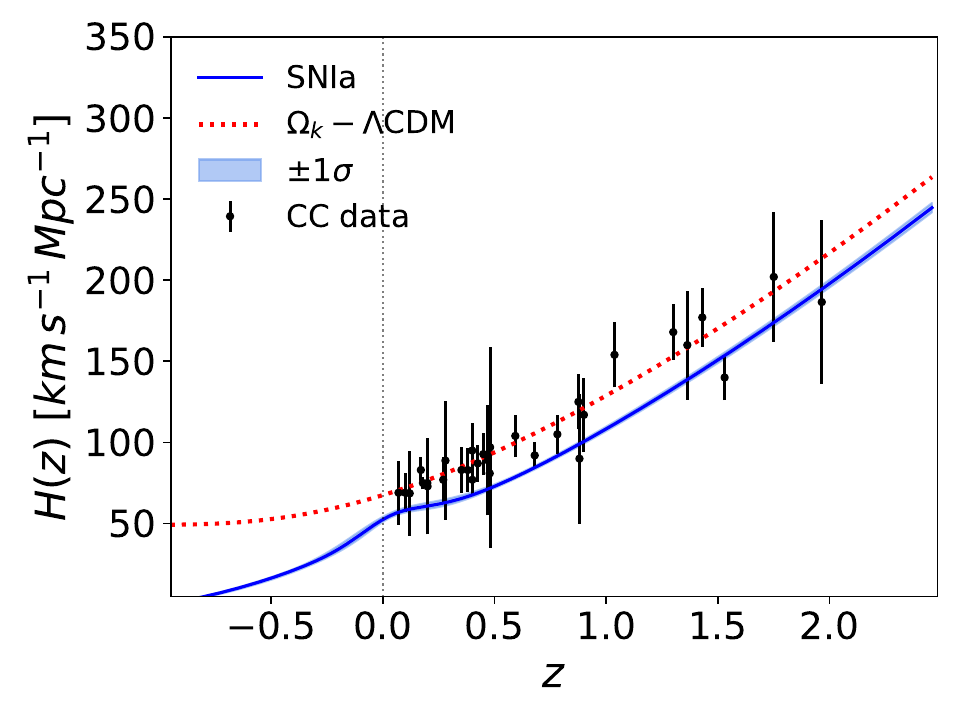}
    \includegraphics[width=0.32\textwidth]{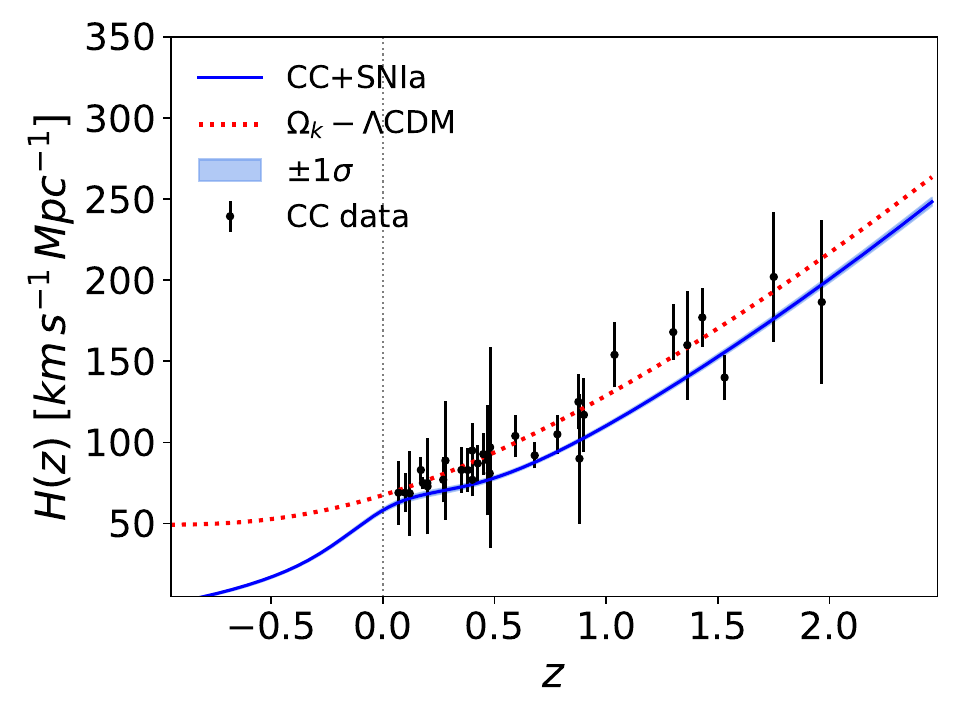}\\
    \includegraphics[width=0.32\textwidth]{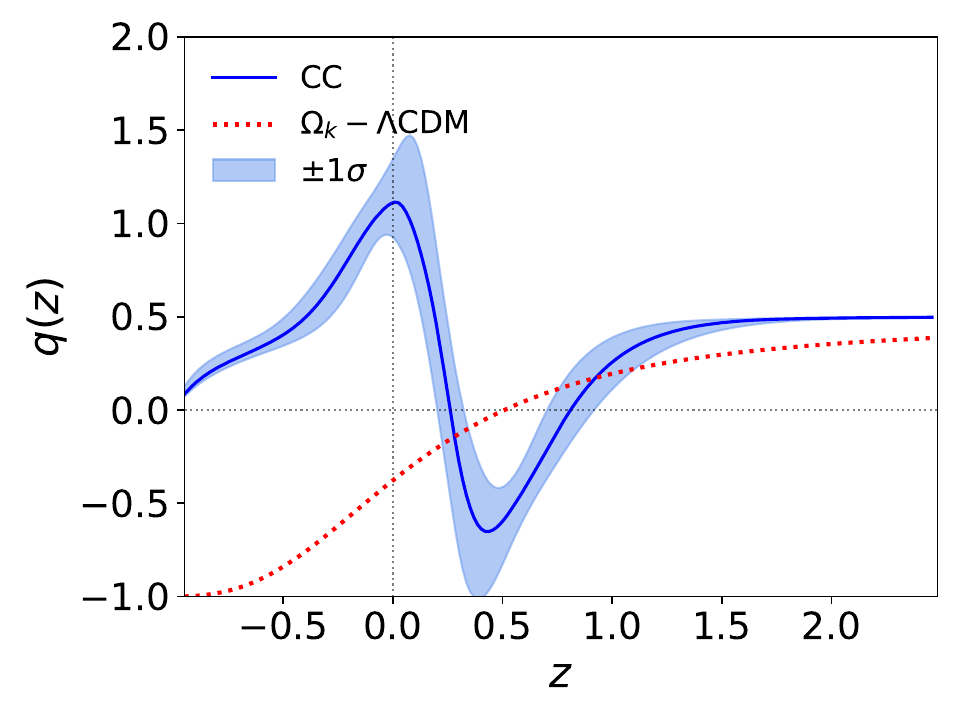}
    \includegraphics[width=0.32\textwidth]{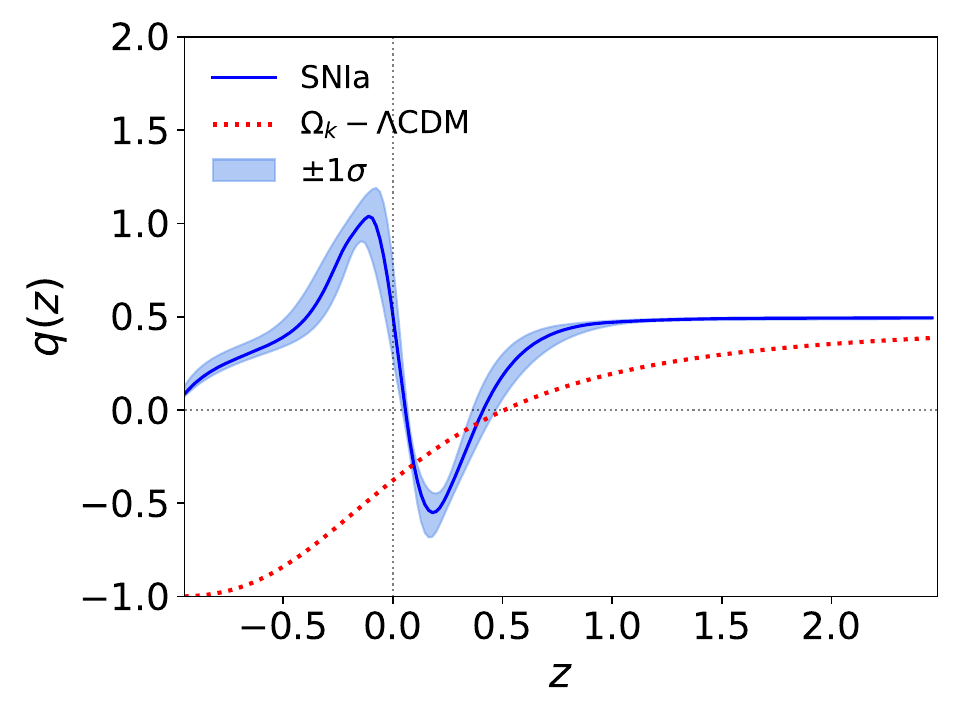}
    \includegraphics[width=0.32\textwidth]{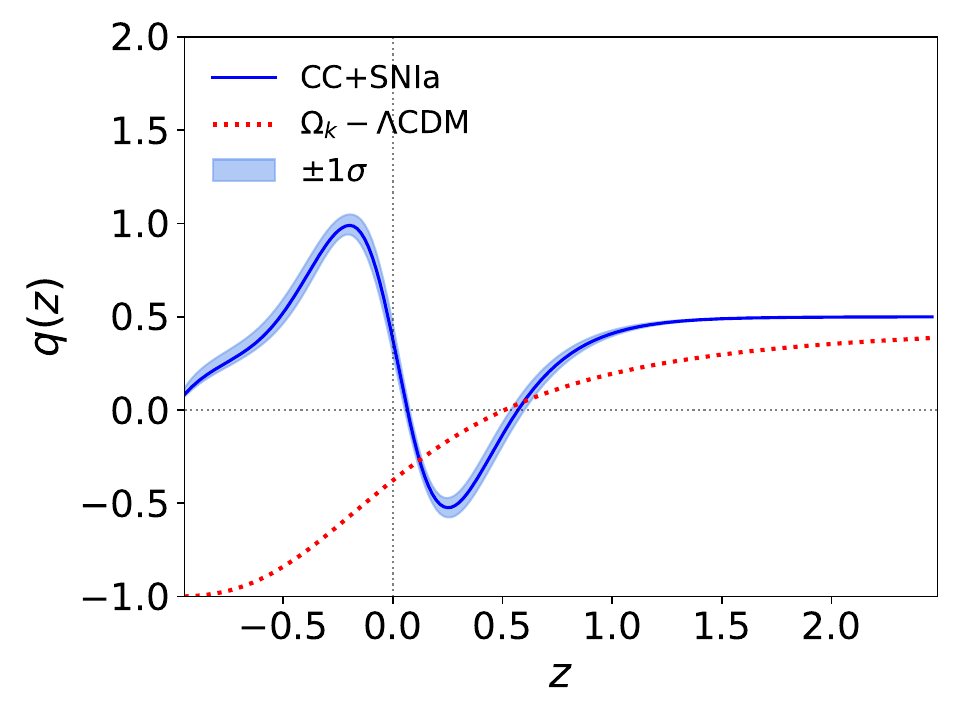}\\  
    \includegraphics[width=0.32\textwidth]{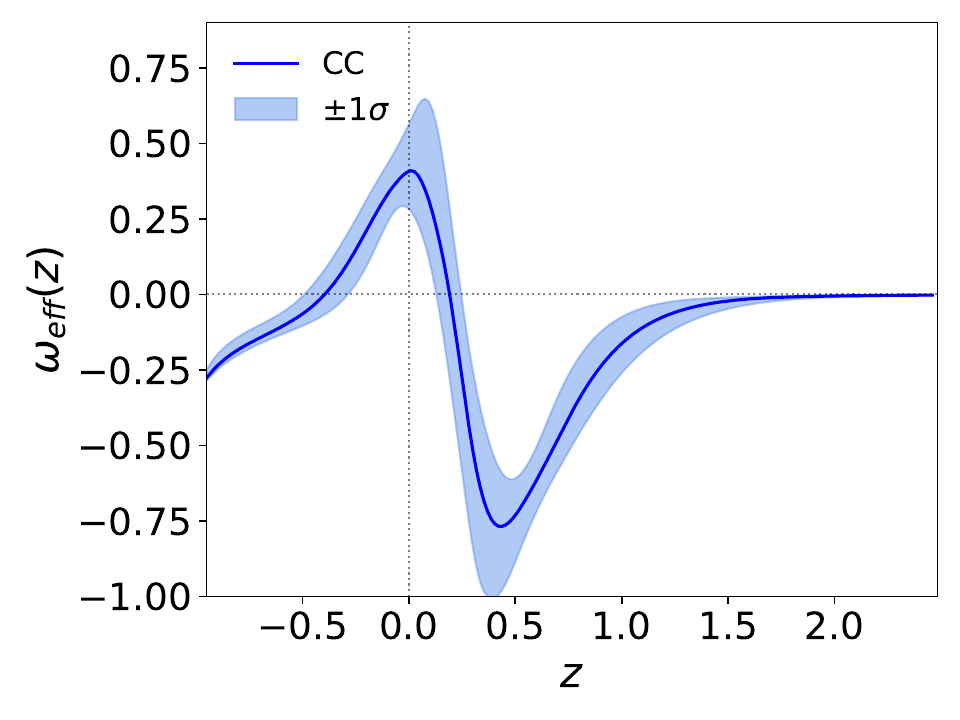}
    \includegraphics[width=0.32\textwidth]{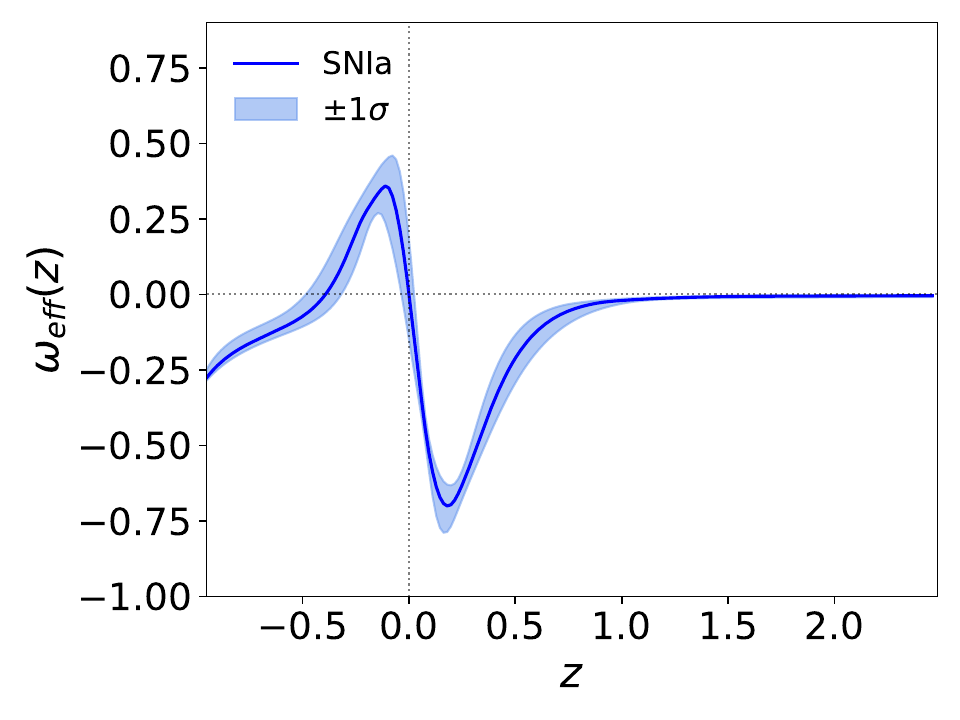}
    \includegraphics[width=0.32\textwidth]{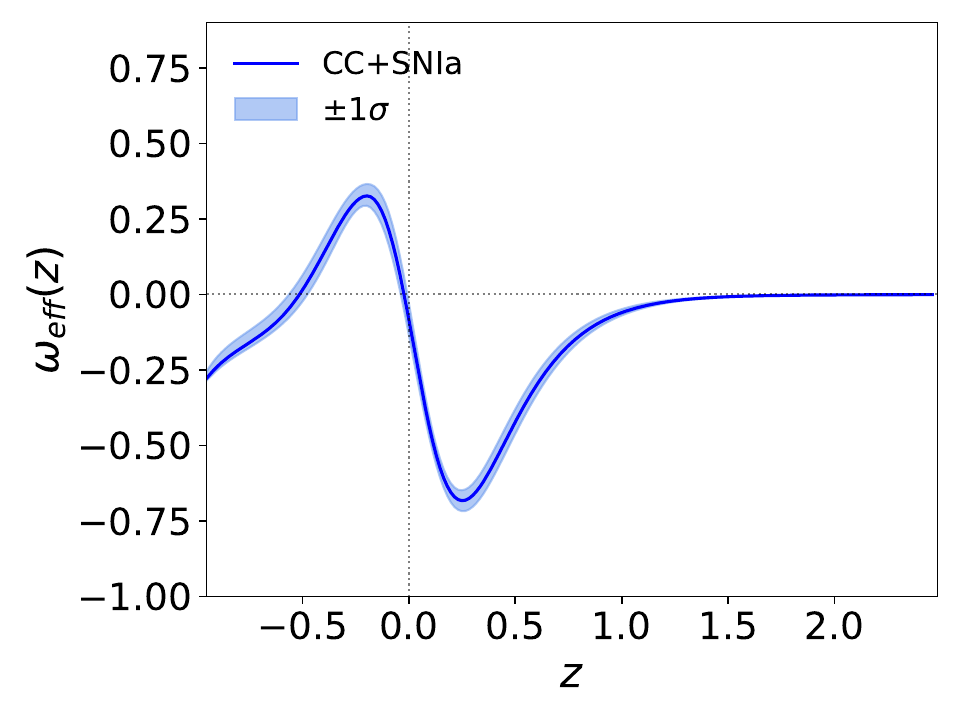}\\ 
    \includegraphics[width=0.32\textwidth]{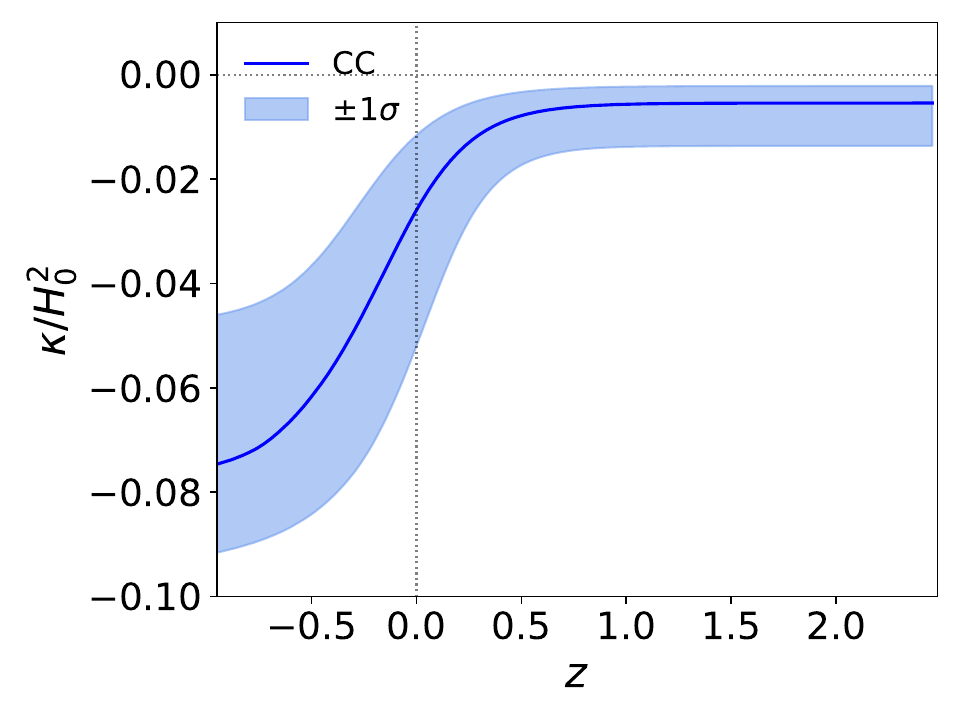}
    \includegraphics[width=0.32\textwidth]{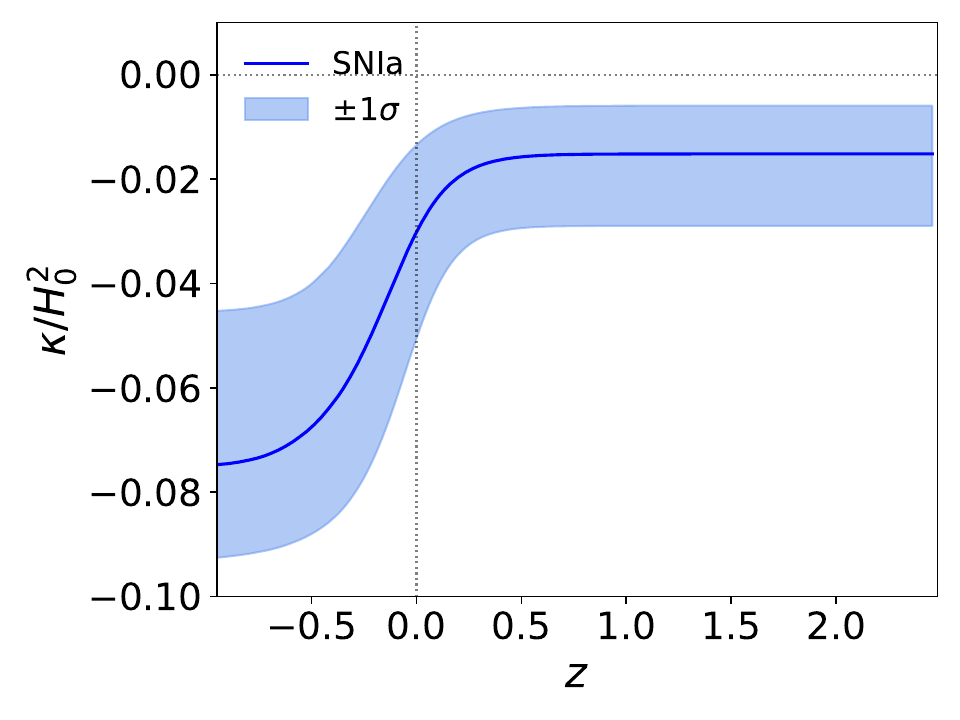}
    \includegraphics[width=0.32\textwidth]{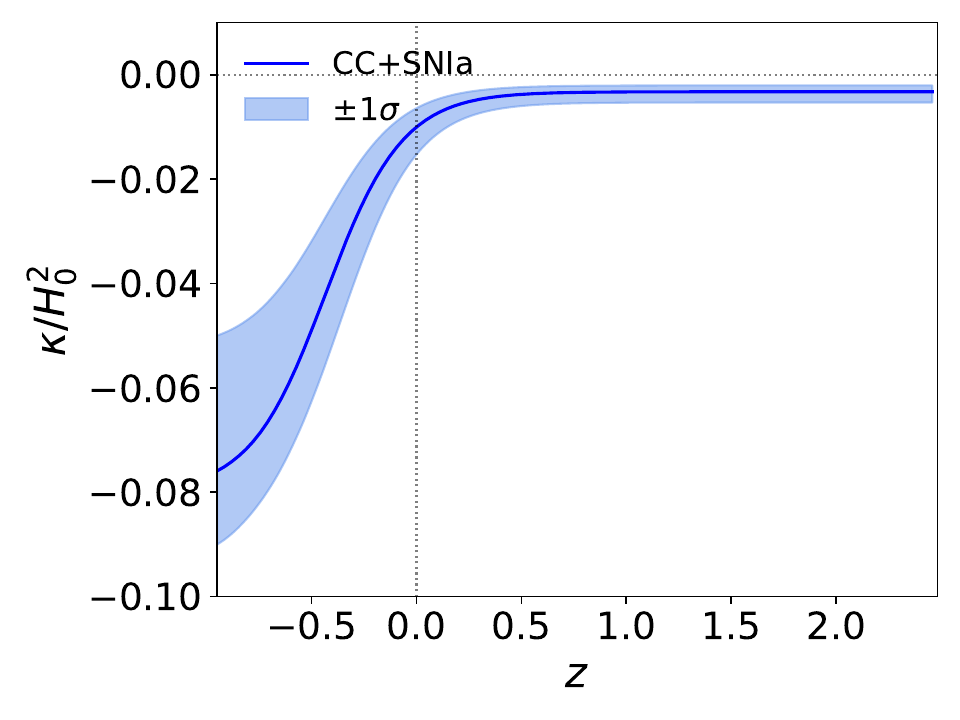}\\ 
    \caption{Reconstruction of the Hubble parameter (first panel), the deceleration parameter (second panel), the effective EoS (third panel) and curvature (fourth panel) for the VC model in the redshift range $-1<z<2.2$ using CC (left), SNIa (middle) and CC+SNIa (right) datasets taking into account Gaussian prior on Planck $H_0$ value. The bands represent $1\sigma$ CL uncertainties around the best-fit values (solid blue line). The standard $\Lambda$CDM model is included in the two first panels as red dashed lines.}
    \label{fig:cosmographyPlanck}
\end{figure*}

Furthermore, Figs. \ref{fig:cosmographySH0ES} and \ref{fig:cosmographyPlanck} present the model reconstruction of the best-fit values of $H(z)$, $q(z)$, $w_{eff}(z)$, and $\kappa(z)/H_0^2$, respectively, in the redshift region $-0.95<z<2.2$ for SH0ES and Planck priors, respectively. For $H(z)$ (the first panel of each Figure), we observe a significant difference from the $\Lambda$CDM model, mainly at redshifts $z < 0$. However, the VC model fit remains consistent with DA observations.
 
Regarding the parameter $\alpha$, the final values constrained by the different probes shown in Table \ref{tab:bf_model}, $\alpha\approx[0.018,-0.132,<-0.1944]$ and $[-0.142,-0.122,<-0.3055]$ for SH0ES and Planck priors, respectively, represent the halfway point of the evolution stage of the curvature, indicating that the transition phase would start at an earlier time. This epoch approximately coincides with the expected $z_{acc}$ at which the acceleration starts to take place according to the $\Lambda$CDM model. 
These features can be seen in the fourth and second panels of Figs. \ref{fig:cosmographySH0ES} and \ref{fig:cosmographyPlanck}, where the function $\mathcal{H}(z)=\kappa/H_0^2$ evolves approximately during the interval $(-1,0.5)$, and the deceleration parameter becomes $q(z)< 0$ around $z\sim[0.4-0.7]$ with a mean around $z\sim 0.5$ as well. This parameter displays oscillations\footnote{This particular behavior also happens in certain DE parameterizations, such as CPL, JBP, among others, as it is shown in \cite{Magana:2017gfs}, where a non-accelerated Universe is considered at $z=0$.} below that transition redshift, $z_T$,  during the curvature transition. In the case of $w_{eff}(z)$, we also observe these oscillations\footnote{Despite the fact that we are comparing $w_{DE}$ from DESI to our $w_{eff}(z)$, it is possible to assess this equality because the effects of $w_{eff}(z)$ at $z\to0$ are only from the DE component.} as previously found in works such as \cite{Zhao:2017} and more recently by the Dark Energy Spectroscopic Instrument (DESI) \cite{DESI:2024mwx} \cite{Escamilla:2024}, with SDSS-eBOSS BAO data.
For the parameter $\gamma$, the CC+SNIa analysis yields a value of $\gamma\approx [5.64, 5.41]$, for SH0ES and Planck priors, respectively. In this case, we assume that this parameter plays the role of triggering the acceleration of the Universe, and its interpretation relates to how smooth the transition is between the curvature values $\Omega_{\kappa}^{1}$ and $\Omega_{\kappa}^{2}$. Specifically, we excluded larger values of $\gamma$ that would not allow for an accelerated transition (as shown in Fig. \ref{fig:bigdelta}).
For instance, the results of the joint analysis CC+SNIa for SH0ES indicate that it is possible to reconstruct the curvature parameter starting the low-phase transition from $\Omega_{\kappa}^{1}>-0.0173$ at $z\approx0.5$ and continuing into the future of the Universe with a stabilization of the curvature at $\Omega_{\kappa}^{2}\lesssim-0.04$, around $z\approx-1$. Notice that it is also possible to infer the value of $\dot{\kappa}$ through Figs. \ref{fig:cosmographySH0ES} and \ref{fig:cosmographyPlanck}, with the maximum of this derivative being $\dot{\kappa}\approx0.1 H_0^3\approx3\times10^{-100}$eV$^3$, which still fulfills the initial assumption of preserving small values of $\dot{\kappa}$.

\begin{table*}
	\centering
	\caption{Marginalized values at 68\% CL around the median value of the posterior constraints for the VC model and $\Omega_\kappa$-$\Lambda$CDM with free curvature parameter. The upper and lower limits are estimated at 90\% CL. Remember that $\Omega_k\equiv-\kappa/H_0^2$ for the standard model and $\Omega_k^i\equiv\kappa^i/H_0^2$ for the VC model.}
	\label{tab:bf_model}
	\begin{tabular}{lccc} 
    \hline
    Parameter  & CC                          & SNIa                        & CC+SNIa \\
    \hline
     \multicolumn{4}{c}{VC, SH0ES prior} \\
 $h$           & $0.736^{+0.010}_{-0.010}$  & $0.731^{+0.010}_{-0.010}$   & $0.743^{+0.009}_{-0.009}$   \\ [0.9ex] 
 $\Omega_{m0}$ & $0.313^{+0.005}_{-0.005}$  & $0.311^{+0.006}_{-0.006}$   & $0.315^{+0.005}_{-0.005}$   \\ [0.9ex] 
 $\Omega_\kappa^1$  & $>-0.0173$ & $>-0.033$ & $>-0.0063$  \\ [0.9ex] 
 $\Omega_\kappa^2$  & $<-0.0398$ & $<-0.0339$ & $<-0.0411$  \\ [0.9ex] 
 $\alpha$      & $0.018^{+0.131}_{-0.247}$  & $-0.132^{+0.088}_{-0.158}$  & $<-0.1944$  \\ [0.9ex] 
 $\gamma$      & $6.805^{+3.490}_{-1.683}$  & $7.280^{+1.883}_{-1.290}$   & $5.642^{+0.582}_{-0.264}$   \\ [0.9ex] 
 $q_0 $        & $1.044^{+0.297}_{-0.266}$  & $0.489^{+0.292}_{-0.225}$   & $0.339^{+0.123}_{-0.096}$   \\ [0.9ex] 
 $\chi^2$      & $17.44$                    & $1990.90$                   & $2008.96$  \\ [0.9ex]
     \multicolumn{4}{c}{VC, Planck prior} \\
 $h$           & $0.678^{+0.004}_{-0.004}$  & $0.676^{+0.004}_{-0.004}$   & $0.681^{+0.004}_{-0.004}$  \\ [0.9ex] 
 $\Omega_{m0}$ & $0.314^{+0.005}_{-0.005}$  & $0.311^{+0.006}_{-0.006}$   & $0.320^{+0.005}_{-0.005}$  \\ [0.9ex] 
 $\Omega_\kappa^1$  & $>-0.0391$ & $>-0.0316$ & $>-0.0117$  \\ [0.9ex] 
 $\Omega_\kappa^2$  & $<-0.0385$ & $<-0.0365$ & $<-0.0437$ \\ [0.9ex] 
 $\alpha$      & $-0.142^{+0.197}_{-0.223}$ & $-0.122^{+0.080}_{-0.151}$  & $<-0.3055$  \\ [0.9ex] 
 $\gamma$      & $5.195^{+1.501}_{-0.755}$  & $7.421^{+1.808}_{-1.341}$   & $5.414^{+0.255}_{-0.129}$  \\ [0.9ex] 
 $q_0 $        & $1.111^{+0.233}_{-0.186}$  & $0.508^{+0.276}_{-0.228}$   & $0.376^{+0.082}_{-0.077}$  \\ [0.9ex] 
$\chi^2$       & $22.56$                    & $1991.86$                   & $2029.30$  \\ [0.9ex] 
 
 \hline
        \multicolumn{4}{c}{$\Omega_{\kappa}$-$\Lambda$CDM, SH0ES prior} \\
 $h$  & $0.727^{+0.010}_{-0.010}$  & $0.730^{+0.010}_{-0.010}$  & $0.709^{+0.009}_{-0.009}$  \\ [0.9ex] 
 $\Omega_{m0}$  & $0.311^{+0.006}_{-0.006}$  & $0.310^{+0.006}_{-0.006}$  & $0.310^{+0.005}_{-0.006}$  \\ [0.9ex] 
 $\Omega_{k}$  & $-0.148^{+0.071}_{-0.067}$  & $0.201^{+0.048}_{-0.047}$  & $0.114^{+0.041}_{-0.040}$  \\ [0.9ex] 
 $q_0 $  & $-0.681^{+0.070}_{-0.066}$  & $-0.333^{+0.046}_{-0.045}$  & $-0.420^{+0.039}_{-0.038}$  \\ [0.9ex] 
$\chi^2$  & $15.93$  & $1992.08$  & $2024.14$  \\ [0.9ex] 
        \multicolumn{4}{c}{$\Omega_{\kappa}$-$\Lambda$CDM, Planck prior} \\
 $h$  & $0.677^{+0.004}_{-0.004}$  & $0.677^{+0.004}_{-0.004}$  & $0.675^{+0.004}_{-0.004}$  \\ [0.9ex] 
 $\Omega_{m0}$  & $0.311^{+0.006}_{-0.006}$  & $0.310^{+0.006}_{-0.006}$  & $0.310^{+0.006}_{-0.006}$  \\ [0.9ex] 
 $\Omega_{k}$  & $0.039^{+0.074}_{-0.072}$  & $0.202^{+0.048}_{-0.047}$  & $0.158^{+0.040}_{-0.040}$  \\ [0.9ex] 
 $q_0 $  & $-0.494^{+0.073}_{-0.071}$  & $-0.333^{+0.046}_{-0.045}$  & $-0.377^{+0.039}_{-0.039}$  \\ [0.9ex] 
$\chi^2$  & $14.54$  & $1992.08$  & $2010.06$  \\ [0.9ex] 
\hline
	\end{tabular}
\end{table*}

Finally, we statistically compare the variable-curvature model and $\Omega_k$-$\Lambda$CDM using two criteria. The first criterion we use is the Akaike Information Criterion (AIC) \cite{AIC:1974, Sugiura:1978} defined as AIC$\equiv\chi^2 + 2k$, being $k$ the number of free parameters and $N$ the size of the sample. The model with the lowest AIC value is the one preferred by the data, and the given model is discriminated against by the following rules. If the difference in AIC value between a given model and the best one ($\Delta$AIC) is less than $4$, then both models are equally supported by the data. For the $\Delta$AIC values in the interval $4<\Delta$AIC$<10$, the data still support the given model but less than the preferred one. For $\Delta$AIC$>10$, the observations do not support the given model. 
According to our results presented in Table \ref{tab:AICBIC}, AIC only suggests a preference to VC model over $\Omega_k$-$\Lambda$CDM for the joint analysis and when SH0ES prior is imposed.

The second criterion we apply, the Bayesian Information Criterion (BIC) \cite{schwarz1978}, gives a stronger penalty than AIC due to the number of degrees of freedom, and is defined as BIC$\equiv\chi^2+k\log(N)$. BIC represents the evidence against a candidate model being the best model (the one with a lower BIC value) and discriminates taking into account the following rules. There is no appreciable evidence against the given model when $\Delta$BIC$<2$. A modest evidence against the candidate model is presented if $2<\Delta$BIC$<6$. The evidence against the candidate model is strong if $6<\Delta$BIC$<10$, and even stronger evidence against is when $\Delta$BIC$>10$. BIC gives a strong evidence against the VC model over when it is compared with the $\Omega_k$-$\Lambda$CDM.

\begin{table*}
    \centering
        \caption{Statistical comparison between the VC model and $\Omega_\kappa$-$\Lambda$CDM using Akaike and Bayesian Information Criteria for both SH0ES prior and Planck prior analysis. $\Delta$AIC and $\Delta$BIC represent the difference between the VC and the $\Omega_\kappa$-$\Lambda$CDM values. Negative values represent a preference to the VC model.} 
    \begin{tabular}{ccccccc}
    \hline
            \multicolumn{7}{c}{SH0ES prior} \\
     Data     & AIC(curvature) & AIC($\Lambda$CDM) & $\Delta$AIC & BIC(curvature) & BIC($\Lambda$CDM) & $\Delta$BIC  \\
     CC       & 29.44          & 21.93             & 7.51        & 38.23	      & 26.33             & 11.91 \\
     SNIa     & 2002.9         & 1998.08           & 4.82        & 2035.53        &	2014.40           & 21.14\\
     CC+SNIa  & 2020.96        & 2030.14           & -9.18       & 2053.60        &	2046.46           & 7.14 \\
     \hline
    \multicolumn{7}{c}{Planck prior} \\
     Data     & AIC(curvature) & AIC($\Lambda$CDM) & $\Delta$AIC & BIC(curvature) & BIC($\Lambda$CDM) & $\Delta$BIC  \\
     CC       & 34.56	       & 20.54	           & 14.02	     & 43.35	& 24.94	& 18.42 \\
     SNIa     & 2003.86	       & 1998.08	       & 5.78	     & 2036.49	& 2014.40 &	22.10 \\
     CC+SNIa  & 2041.3	       & 2016.06	       & 25.24	     & 2074.04	& 2032.43&	41.61 \\
\hline
    \end{tabular}
    \label{tab:AICBIC}
\end{table*}

 \section{Discussion and conclusions} \label{sec:DisandCon}

In this paper, we study the evolution of the Universe under a variable curvature (VC) scenario, assuming the validity of GR. For this case, we restrict the form of the curvature transition by imposing the shape given in Eq. \eqref{AnalyticCurv}. Additionally, we expect that the curvature variation will be able to produce a local Universe acceleration without the need for a DE component, such as the cosmological constant considered in the current standard cosmological model. Interestingly, in the scenario presented in this work, this behavior is possible because the curvature is now coupled with the acceleration equation, unlike the standard case in which the curvature term does not appear. According to our constraints, the $\gamma$ parameter seems to act as the entity responsible for the current acceleration of our Universe.

In order to analyze the best prior regions for the likelihood Bayesian analysis, we first performed a theoretical study to determine the values needed to obtain an accelerated Universe with a non-significant violation of the cosmological principle. In this case, our free parameters were $\Omega_\kappa^{1}$ and $\Omega_\kappa^{2}$, that are related to the $\alpha$ parameter, which represents the curvature transition in terms of redshift, and $\gamma$, which is associated with the parameter that generates the Universe acceleration and softens the $\mathcal{H}(z)$ function. Additionally, $\Omega_{m0}$ and $h$ denote the matter density and dimensionless Hubble constant, respectively. According to our results detailed in Appendix \ref{SubThe}, these initial values correspond to $\Omega_{m0}=0.33$, $\alpha=-0.09$, $\gamma=10$, $\Omega_\kappa^1=-0.0007$, $\Omega_\kappa^2=-0.004$, while also maintaining, to a first approximation, the Friedmann constriction as $E(z=0)=1.02$.

Thus, our next step was to implement the complete likelihood MCMC analysis using CC, SNIa and a joint analysis that incorporates both data samples, obtaining the marginalized values shown in Table \ref{tab:bf_model} and the 1D posterior distributions and 2D contours at levels $1\sigma$ (inner) and $2\sigma$ (outer) in Figs. \ref{fig:contoursSH0ES} and \ref{fig:contoursPlanck}. Notice that for the matter-energy density case, we obtain values in agreement with those expected in the standard model, while in this new proposed model, the curvature density parameter varies between $\Omega_{\kappa}^{1}>-0.0063$ for the CC+SNIa analysis with SH0ES prior ($>-0.0117$ with Planck prior) and $\Omega_{\kappa}^{2}<-0.0411$ with SH0ES prior ($<-0.0437$ with Planck prior). This indicates a change in curvature during the interval of epochs $z\sim(-1,0.5)$, with an average value for the transition at $\alpha<-0.1944$ for SH0ES prior ($\alpha<-0.3055$ for Planck prior). This produces a transition to an accelerated Universe at $z_T =0.605^{+0.106}_{-0.135}$ (CC), $0.415^{+0.058}_{-0.052}$ (SNIa), and $0.522^{+0.035}_{-0.041}$ (CC+SNIa) using SH0ES prior and $z_T =0.787^{+0.117}_{-0.154}$ (CC), $0.412^{+0.057}_{-0.050}$ (SNIa), and $0.570^{+0.029}_{-0.033}$ (CC+SNIa) using Planck prior, all of them consistent with the $\Lambda$CDM model's value of $z_{acc}\sim0.6$ with a flat curvature. 

Moreover, Figs. \ref{fig:cosmographySH0ES} and \ref{fig:cosmographyPlanck} show coincidence at $1\sigma$ with the standard model in the evolution of $H(z)$, however at $z=-1$, that is, $a\to\infty$, the Universe tends to $H(z)=0$, differentiating from the $\Lambda$CDM standard model where $H(z)=H_0\sqrt{\Omega_{0\Lambda}}$.  Meanwhile, for the deceleration parameter, we see a transition to an accelerated Universe close to where it is expected. The main difference lies in the fact that, for $z=0$, the VC model predicts a re-transition to a decelerated stage and exhibits an oscillatory trend. This behavior is also observed in several DE parameterizations like CPL, JBP, among others \cite{Chevallier:2000qy,Linder:2002et,Jassal:2004ej,Castillo_Santos_2023}. We need to mention that our model tends to decelerate when the $q(z)$ function is evaluated at $z=0$ (for the SNIa and CC+SNIa results), similar to various DE parameterizations such as Jassal-Bagla-Padmanabhan (JBP), Barboza-Alcaniz (BA), among others \citep{Magana:2017gfs}. Furthermore, we observe that the effective EoS $w_{eff}$ also presents an oscillatory behavior, an effect that has been widely discussed in recent results from BOSS and eBOSS data \cite{Zhao:2017, Escamilla:2024} and especially by the DESI collaboration \cite{DESI:2024mwx}. 

In the following, we address the advantages of this VC model over the current standard cosmology, as well as discuss future analyses to clarify the possible processes in charge of driving the low-phase curvature transition. 

First, the acceleration in the late times of the Universe is caused by a small change in curvature, characterized by the parameter $\gamma$, which produces a softening of the step function and a transition that preserves $\dot\kappa$ close to zero.  This change in curvature could be caused by some rearrangement of the fluids that permeate the Universe; this change is also small (flat) but has a tendency towards a hyperbolic Universe. Notice that the Planck satellite observes a value of curvature of $\Omega_\kappa^{\Lambda\rm{CDM}}=0.001\pm0.002$ \cite{Planck:2018}, and similar values were also obtained by previous Wilkinson Microwave Anisotropy Probe (WMAP) observations \cite{WMAP:2007ApJ}, compatible with our results in the late stages of the Universe evolution. Furthermore, this negative region in the curvature density parameter allows for an acceleration stage of the Universe without the need for a cosmological constant, as shown in Figs. \ref{fig:cosmographySH0ES} and \ref{fig:cosmographyPlanck}.

Of course, a curvature with temporal dependence can not be understood with the traditional FLRW equations, instead an extension must be implemented always within the framework of GR, which represents a clear advantage because it allows us to avoid dealing with any type of GR extensions (like in \cite{Granda-2020,Goswami-2023}). However, we need to be cautious about the proposed curvature shape, that should always be under the restriction given by $\dot{\kappa}\approx0$, otherwise we would be entering the scenario of a Stephani Universe \cite{Stephani:2003tm}. Remember that we also considered a step function and its derivative counterpart, the Dirac delta function, $\delta(z)$, as an appropriate candidate to fulfill $\dot{\kappa}\approx0$. However, we observed that in this case it is not possible to obtain an accelerated stage for the expected redshift range close to our local Universe. Therefore, we would not be considering an inhomogeneous scenario, but rather a slight violation of the FLRW framework during a low-phase transition with very similar initial and final curvature values, in order to explore the possibilities of small deviations from the perfect cosmological principle.

Besides, this model predicts a decelerated stage at $z=0$ in comparison with the standard model. Still, this result is totally consistent with observations, as it is not possible to assert whether our Universe is in an accelerated or decelerated phase in the present epoch at $z=0$ (as discussed by \cite{Shapiro-2006}). Yet, the accelerated behavior comes from an extrapolation of the fit to the SNIa data of the $\Lambda$CDM model. In this context, our model shows an interesting oscillation, which is characteristic of many parameterizations of DE. However, in this case, the oscillation is a natural outcome of this model. Specifically, we see that we would not need a cosmological constant in the usual form of DE to justify the Universe acceleration but rather a curvature transition that can simulate some kind of a dynamical DE with an oscillating shape (see \cite{Zhao:2017,DESI:2024mwx, Escamilla:2024}).

Finally, one of the possible future challenges is to investigate this perturbative part by adding the variation in curvature into the field equations and their corresponding couplings with the Boltzmann equations. Moreover, regarding the continuity equation, our calculations show a standard behavior without any kind of coupling with curvature, except for the transition phase at low redshifts (as discussed in Appendix \ref{SubThe}). Thus, the coupling of fluids with curvature and its addition in the Friedmann equation need to be further studied, mainly using observations without any model dependence, specially on the standard model. Furthermore, a perturbative analysis of the VC model is required to investigate how this coupling could impact the growth of structures.

We need to be clear that we do not have a definitive answer regarding why a change of curvature occurs in the late times of the Universe evolution. This behavior according to our calculations could be produced by a rearrangement of the galactic structure or due to other more fundamental processes in the structure of the space-time, such as perturbations in the density field or a transition phase governed by the curvature density instead of a vacuum energy with its typical exponential growth during the inflation era. Additionally, we understand that the model contains too many free parameters, and so, we may need to eliminate some of them in a future, more in-depth analysis. 
Moreover, we saw that the Hubble constant could also be slightly modified from its current value, derived from the change in the value of $E(z=0)$ (see Section \ref{sec:Res}). This could suggest that further estimations with additional constraints may be useful in determining whether this VC model can help in alleviating the Hubble tension, one of the current challenges facing the $\Lambda$CDM model (see \cite{LambdaCDMchallenges}).

\begin{acknowledgments}
 A.E.G. acknowledges support from project ANID Fondecyt Postdoctorado with grant number 3230554. M.A.G.-A. acknowledges support from c\'atedra Marcos Moshinsky, Universidad Iberoamericana for the support with the National Research System (SNI) grant and the project 0056 from Universidad Iberoamericana: Nuestro Universo en Aceleraci\'on, energ\'ia oscura o modificaciones a la relatividad general. The numerical analysis was also carried out by {\it Numerical Integration for Cosmological Theory and Experiments in High-energy Astrophysics} (Nicte Ha) cluster at IBERO University, acquired through c\'atedra MM support. A.H.A. thanks the support from Luis Aguilar, Alejandro de Le\'on, Carlos Flores, and Jair Garc\'ia of the Laboratorio Nacional de Visualizaci\'on Cient\'ifica Avanzada. M.A.G.-A, A.H.A, J.M. and V.M. acknowledge partial support from project ANID Vinculaci\'on Internacional FOVI220144.
 \end{acknowledgments}
 
\begin{appendix}

\section{Einstein field equations}

The Einstein field equations provide us with $tt$, $rr$ and $\theta\theta$ equations written as

\begin{eqnarray}
&&2r^4a\kappa\ddot{\kappa}-3r^4a\dot{\kappa}^2-12r^4\kappa^2\ddot{a}+4r^4\kappa\dot{\kappa}\dot{a}-2r^2a\ddot{\kappa}\nonumber\\&&+24r^2\kappa\ddot{a}-4r^2\dot{a}\dot{\kappa}-12\ddot{a}=16\pi Ga(3p+\rho)\times \nonumber\\&&(r^2\kappa-1)^2, \label{uno}
\end{eqnarray}

\begin{eqnarray}
&&2r^4a\kappa\ddot{\kappa}-3r^4a\dot{\kappa}^2-4r^4\kappa^2\ddot{a}+8r^4\kappa\dot{\kappa}\dot{a}-8r^4a^{-1}\kappa^3\nonumber\\&&-8r^4\kappa^2a^{-1}\dot{a}^2-2r^2a\ddot{\kappa}+8r^2\kappa\ddot{a}-8r^2\dot{a}\dot{\kappa}+16a^{-1}r^2\kappa^2\nonumber\\&&+16r^2a^{-1}\kappa\dot{a}^2-4\ddot{a}-8a^{-1}\kappa-8a^{-1}\dot{a}^2\nonumber\\&&=-16\pi G(\rho-p)a(r^2\kappa-1)^2, \label{dos}
\end{eqnarray}

\begin{eqnarray}
    &&2r^2\kappa\ddot{a}-r^2\dot{a}\dot{\kappa}+4r^2a^{-1}\kappa^2+4r^2\kappa a^{-1}\dot{a}^2-2\ddot{a}\nonumber\\&&-4a^{-1}\kappa-4a^{-1}\dot{a}^2=8\pi G(\rho-p)a(r^2\kappa-1), \label{tres}
\end{eqnarray}
and $tr$ component is $r\dot{\kappa}(1-r^2\kappa)^{-1}$, which must be equal to 0 for traditional FLRW metrics.

\section{A theoretical analysis of the free parameters} \label{SubThe}

We begin a test for the preliminary inspection of the VC model by studying a set of preferred values for the free parameters. The aim is to restrict the variance in the free parameters as much as possible before applying the likelihood analysis with the data.

Our first validation step is to propose small values for $\Omega_{\kappa}^i$ and $\gamma$ in order to partially maintain the cosmological principle or, alternatively, considering $\gamma\to\infty$ 
to obtain a steep and fast transition that mimics a Dirac Delta function between two close values of $\Omega_{\kappa}^i$. Both approaches were carried out using the Friedmann dimensionless equation, along with its corresponding substitution in the deceleration parameter equation, which was calculated numerically, and to accelerate the Universe through the parameter $\gamma$. 

After exploring different combinations, starting from the common density parameters of $\Lambda$CDM, we arrive at a set of proposed theoretical parameters: $\Omega_{m0}=0.33$, $\alpha=-0.09$, $\gamma=10$, $\Omega_{\kappa}^1=-0.0007$, $\Omega_{\kappa}^2=-0.004$, $E(0)=1.02$. Through a numerical integration, we obtain $t_{age}=1.4\times10^{10}$yrs, the age of the Universe. Here, we inspected possible values of curvature close to 0, allowing for small changes around this flat prior in both positive and negative Universes. We found that, in order to retrieve an accelerating Universe at local times and to  obtain values with a non-substantial deviation from the premise $E(0)=1$, we need to keep the $\gamma$ parameter small, thereby having a model in which only a low-phase transition is allowed. Furthermore, we first tried to look for values $\alpha$ close to $z\sim 0.6$, however, since the transition demanded small values $\gamma$ (ranging from 1 to 100, with good behavior observed for values $\lesssim10$), the obtained value was $\alpha=-0.09$. This value represents an average location of the total transition, indicating that the starting point of the transition happens earlier (interestingly, at $z\sim0.6$). The rest of the combinations in which the $\gamma$ parameter is larger encountered problems in fitting the expansion of the Universe in a way consistent with real data, either accelerating the Universe at late times or fulfilling the requirement of $E(z=0)\simeq1$. 
To show this, in Fig. \ref{fig:bigdelta} we explore the case where $\gamma=100$, reproducing almost a step function\footnote{Notice that when $\gamma\to\infty$ in Eq. \eqref{AnalyticCurv}, the step function is recovered.} and clearly, for this case, the VC model cannot reach an accelerated stage. As in the second case, we keep the same values for the parameters, but now considering $\gamma=10$. In this case, the evolution of the deceleration parameter transits from decelerated to accelerated (see also Fig. \ref{fig:bigdelta}). Thus, a lower value of $\gamma$ is needed to keep the condition $\dot{\kappa}\approx0$ and with this, an accelerated phase, as discussed in the text and corroborated by our theoretical constraints.

From this theoretical analysis, we also observe that only a negative value of $\kappa$ (hyperbolic geometry)\footnote{By definition, in the common literature, the sign of $\kappa$ is the opposite of that of $\Omega_{\kappa}$. However, by our own choice, we have used the same signs for both $\kappa$ and $\Omega_{\kappa}^{i}$ as seen in Eq. \eqref{kappa_function}.} is able to reproduce the acceleration predicted by the $\Lambda$CDM cosmology. For this reason, the possibility of a closed Universe is excluded from our later analysis. Thus, for the calculation of the luminosity distance for this VC model, we will apply the case of a hyperbolic Universe, which does not exclude the flat case, because in the limit when $\Omega_\kappa=0$ in the hyperbolic definition, we recover the flat case as follows
 
 \begin{eqnarray}
    &&\lim_{\Omega_\kappa\to0}(1+z)\frac{c}{H_0\sqrt{\Omega_\kappa}} \sinh\left[\sqrt{\Omega_\kappa}\int_0^z \frac{dz^{\prime}}{E(z^{\prime})}\right]=\nonumber\\&&(1+z)\frac{c}{H_0}\int_0^z \frac{dz^{\prime}}{E(z^{\prime})}, \label{limithypflat}
\end{eqnarray} \\
converging to the $d_{L}$ definition for a flat Universe.

Finally, we perform the test corresponding to the evolution of the matter density given by Eq. \eqref{SolCE}. In principle, we can see an initial coupling of this ratio to the curvature, which also depends on an integration constant defined as $\kappa_{0}=\xi H_{0}^{2}$. However, for a perfect fluid, it is required that this perturbative curvature term approaches the unity, as it is expected that the curvature remains constant throughout the Universe evolution. To analyze the impact of this additional term on the density parameter, we realize that the $\xi$ term must be equal to the initial curvature, $\Omega_{\kappa}^{1}$, and thus, by substituting $\omega=0$ (as expected for the matter component), we arrive at the evolution form of the matter density for the VC model, as shown in Fig. \ref{fig:densityevol}. It can be seen that the model behaves properly (i.e. without any deviation from the traditional $\Lambda$CDM model behavior) except for a very small redshift region corresponding to the transition between curvatures, moderated by $\alpha$ and $\gamma$ parameters. These parameters were taken to be the same as for the previous theoretical exploration, providing a starting point for the curvature transition at $z_{T}\sim0.4$. We can conclude that the matter density evolution is only affected in the very local Universe (in fact, the ratio at $z=0$ in this theoretical exploration should be multiplied by a factor of 1.5). Therefore, we consider that, in general, there is no coupling between the matter component and the curvature for most of the history of the Universe, and we can safely approximate Eq. \eqref{SolCE} as
\begin{eqnarray}
    \rho_m\approx\rho_{m0}a^{-3},\label{rho_m_approx}
\end{eqnarray}
which is the standard form for the matter evolution.

\renewcommand\thefigure{\thesection.\arabic{figure}} 
\setcounter{figure}{0}


\begin{figure*}
    \centering
    \includegraphics[width=0.7\textwidth]{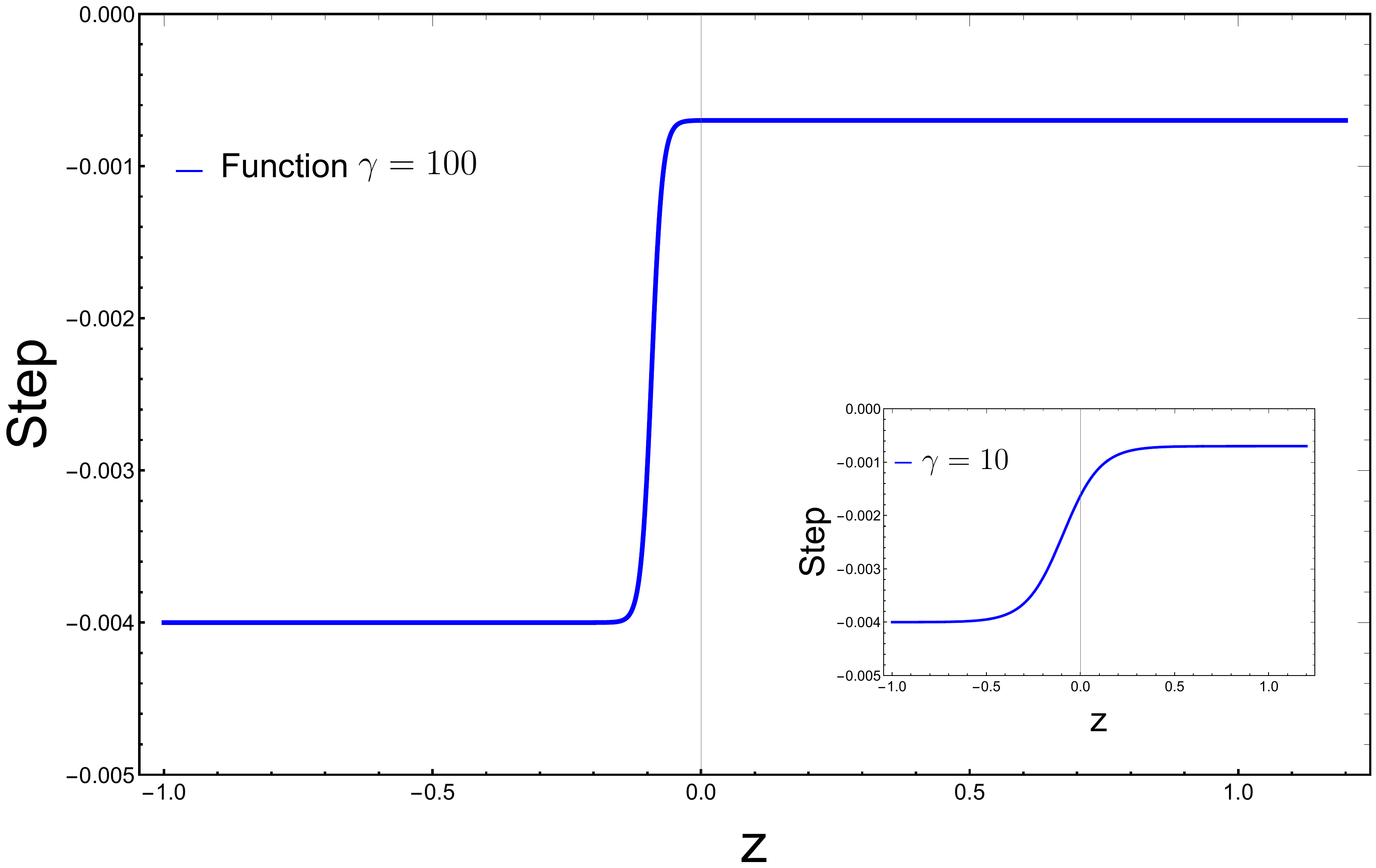} \\
    \hspace*{0.3cm}\includegraphics[width=0.685\textwidth]{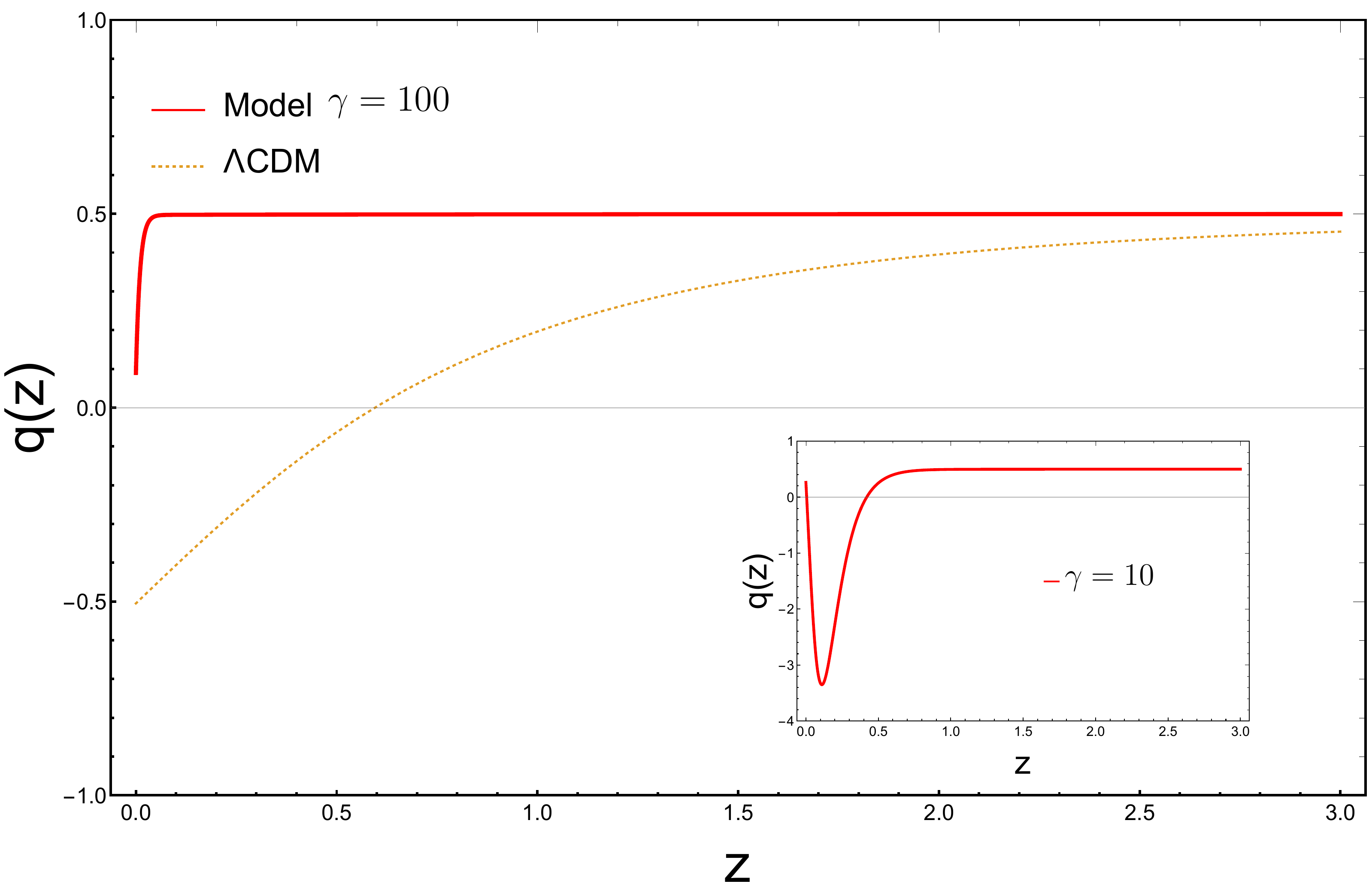}\\
    \caption{Theoretical exploration of the VC model considering the cases of $\gamma=100$ and $\gamma=10$, with the other parameters as shown in Appendix \ref{SubThe}. Upper: Step function $\mathcal{H}(z)$. Lower: Deceleration parameter $q(z)$ including the comparison to the $\Lambda$CDM model (yellow dashed line). Notice how we reproduce a step function, as expected, with a steeper transition between curvatures in the case of $\gamma=100$ and $q(z)$ for the VC model, for which the accelerated state is only reached at local $z$ for $\gamma=10$, contrary to the case of $\gamma=100$. Additionally, both cases differ in the shape of $q(z)$ compared to the standard cosmological model.}
    \label{fig:bigdelta}
\end{figure*}

\begin{figure*}
    \centering
    \includegraphics[width=0.65\textwidth]{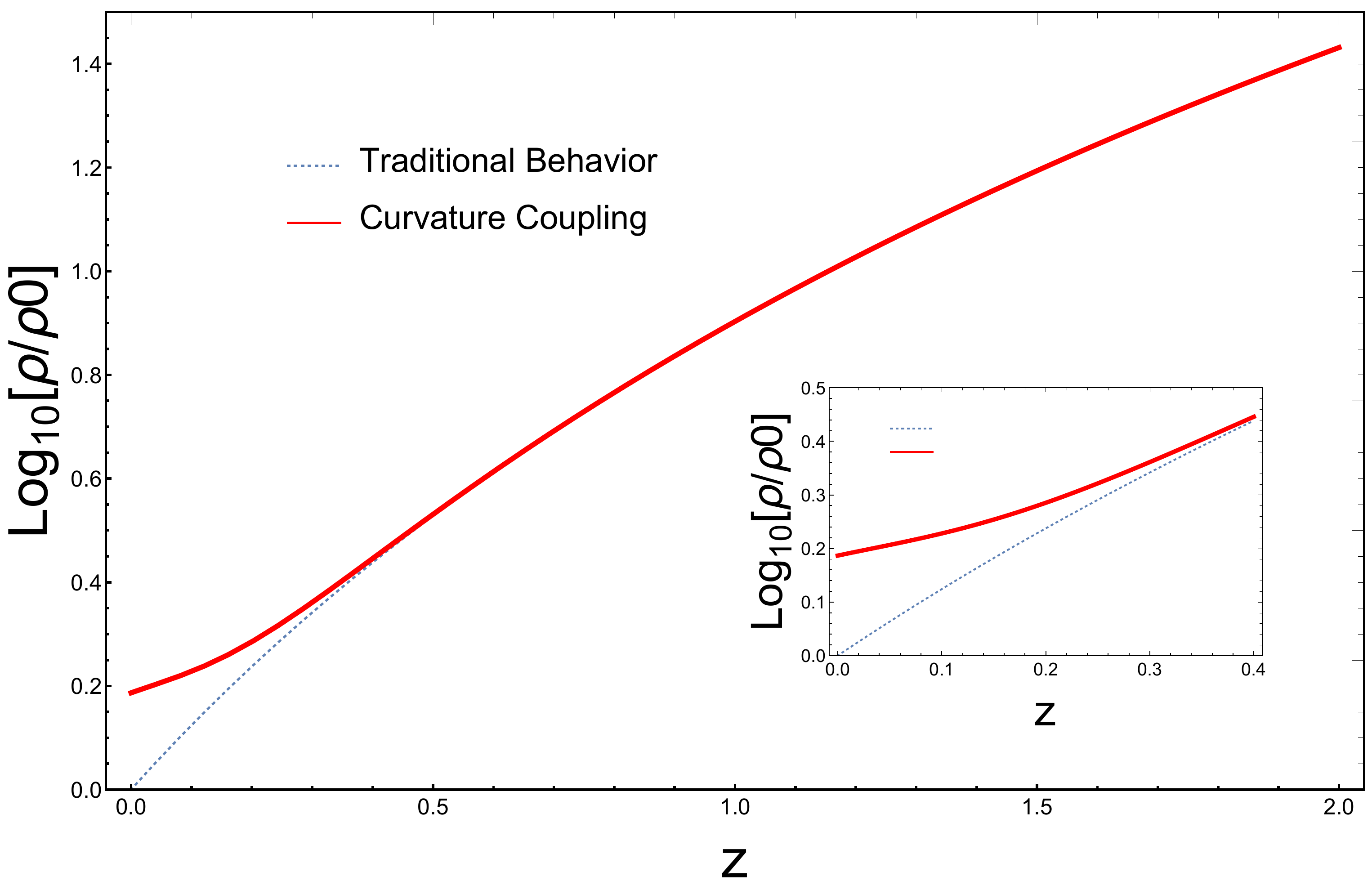}\\
    \caption{Logarithmic matter density evolution ($w=0$) with the selected theoretical free parameters (see text). It shows the comparison between this value coupled with $\mathcal{H}(z)$ (solid red line) and without the coupling (traditional evolution, blue dashed line). The main differences in the energy density ratio are observed near the curvature transition phase around $z=\alpha$ (see the small zoomed-in square), where this ratio $\rho/\rho_{0}$ now multiplied by a factor $\sim 1.5$ at local time ($z=0$).}
    \label{fig:densityevol}
\end{figure*}
\end{appendix}

\clearpage
\bibliography{references}


\end{document}